# A Critical Appraisal and Motivated Intruder Test of A Search Dataset Anonymised for the Purposes of the EU Digital Markets Act

Ghislain d'Entremont*[1,2], Lisa Pilgram*[1,2,4], Sophie Stalla-Bourdillon*[3], Renee Gil[1,2], Lucy Mosquera[1], Olga Vovk[2], Khaled El Emam[1,2,4]

[1]Woodway Assurance, Ontario, Canada;

[2]Children's Hospital of Eastern Ontario Research Institute, Ontario, Canada;

[3]Brussels Privacy Hub, LSTS, Vrije Universiteit Brussel, Belgium

[4]School of Epidemiology and Public Health, University of Ottawa, Ontario, Canada;

*equal contributions

# Abstract

Article 6(11) of the Digital Markets Act (DMA) requires gatekeepers to give third-party online search engines access, on fair, reasonable and non-discriminatory terms, to ranking, query, click and view data generated on its online search engines. In that context the European Commission then specified the anonymisation measures that Alphabet must apply before sharing the data. In this study, we perform a critical analysis of the Commission's anonymisation measures to determine the extent to which they follow the European Data Protection Board (EPDB) Guidelines for a contextual approach to anonymisation, and contemporary good anonymisation practices. We also perform a motivated intruder test on a sample of Google searches to determine if we could re-identify individuals and assess the re-identification risk.

Our critical analysis notes that the Commission's anonymisation criteria only cover the end-users of the searches and do not cover individuals that may be the subject of the search queries, as acknowledged by the Commission itself. This means that there may be many individuals with personal information in the queries even if Alphabet implements all of the Commission's anonymisation measures. Furthermore, the Commission's anonymisation measures and tests do not explicitly cover inferences, which is one of the three criteria in the EDPB Guidelines. The controls and safeguards stipulated by the Commission are consistent with those recommended by other regulators and reflect good contemporary practices for managing the residual risk in data that had technical anonymisation transformations applied to them.

The main motivated intruder test targeted 98,501 individuals across the six covered countries. The process involved identifying real and named individuals who engaged in specific activities, such as conferences or sports events, developing a profile on them using public sources of information, and then locating their search records. The profile included information such as name, country and NUTS 3 region, language of a public biography, a date window of up to one week taken from the event programme, professional position, employment history, education, languages, skills, hometown, hobbies and travel details such as conference attendance. The suspected re-identifications were verified by the Google team. This resulted in a re-identification rate of 0.094%. The average effort for every verified match was 3 hours (excluding Google verification time, and Gemini costs; and 1.6 hours further excluding planning and strategy efforts), which is equivalent to €105 in cost. These are low costs and would not realistically act as a deterrent for a re-identification attack. Comparing the 0.094% targeted population re-identification rate that we obtained to our pre-defined threshold, the 0.094% is below that insignificant risk 0.47% threshold. The re-identification rate varied by country and region. Despite the variation, the risk was below the threshold across all countries. One region did have a risk value above the threshold.

The determination of the 0.47% threshold is important. We started from the most conservative precedents in contemporary standards and regulatory guidance where quantitative thresholds were provided. Then adjusting that threshold for a priori data quality and verification rates, and assuming five independent re-identification attack teams were engaged, we worked backwards to determine the data vulnerability threshold after taking into account the controls and safeguards. This is the threshold for any one of the teams, such as ours.

A number of different sensitivity analyses were performed on the assumptions behind the calculation of the threshold, and how uncertain matches were handled in the calculation of the re-identification rate. Under some of these scenarios the identity disclosure re-identification risk was above the threshold. However, these high risk scenarios make strong assumptions, such as assuming that all uncertain matches are actually correct matches. Therefore, we do not draw strong conclusions from them.

All of the matches were verified rather than assumed, and the protocol for the motivated intruder test was approved by our research ethics board. The former is important because unverified matches can be optimistically misleading, and the latter did impose constraints on how the motivated intruder test was performed.

The main conclusion is that the Commission's measures do not clearly cover all of the criteria of the EDPB Guidelines, that the focus only on end-users will still leak the personal information of other individuals covered in the queries, and that the controls and safeguards stipulated by the Commission are consistent with the contextual approach in general and contemporary risk management practices, and such controls and safeguards are necessary for managing residual risk after transforming the queries. From the perspective of identity disclosure only, applying the framing of a contextual approach and within the limitations of our attacks, we found that the re-identification risk would be considered insignificant.

Our motivated intruder testing assumed (and did not directly test) that the Named Entity Recognition tools used to detect direct identifiers and then handle them (i.e., suppression of search queries where the direct identifiers are not in the allow-list) had sufficiently high recall. The smallest defensible recall would be 0.88, and the suggested recall is 0.976 which would account for data errors and verification. The former recall value is achievable with todays NLP methods, and the latter would be more challenging to meet in practice. Even then, there would remain direct identifiers not in the allow-list in the Search Dataset as the recall of these tools is not going to be perfect.

The motivated intruder test targeted a specific population of individuals (such as those who attended events) and the results may not apply to other sub-populations within the covered countries.

Our attempts to target individuals with rare characteristics in the search queries and matching them with real named individuals using only publicly available information had a low yield. A more focused effort on these types of attacks may improve their re-identification rate. Arguably, the additional controls and safeguards stipulated by the Commission are intended to manage such residual risks.

# 1. Introduction

Article 6(11) of the Digital Markets Act (DMA) requires gatekeepers to give third-party online search engines access, on fair, reasonable and non-discriminatory terms, to ranking, query, click and view data generated on its online search engines [1]. This obligation is intended to lower the barrier to entry for new search engines created by the large scale of query logs incumbent engines have access to.

On 27 January 2026, the European Commission opened specification proceedings to help Alphabet comply with its data sharing obligations. The proceedings aimed to set out the measures Alphabet must put in place so that eligible beneficiaries, including organisations developing AI chatbots with search functionalities, can effectively access that data. In its implementing decision of 16 July 2026, adopted after a public consultation on provisional measures published in April [2], the European Commission then specified the anonymisation measures that Alphabet must apply before sharing the data [3]. The anonymisation standard applied by the European Commission determines both the privacy exposure of end users and the utility of the data released to competitors.

In this paper, we contribute to the discussion on the effectiveness of anonymisation measures for search query data. We examine the measures specified by the Commission for the query dataset shared by Alphabet, combining a theoretical analysis with a motivated intruder test on the data provided by Alphabet. We refer to this dataset as the *Search Dataset*. The motivated intruder test serves as an independent empirical assessment of the residual re-identification risk in this dataset.

In this assessment, we consider the combined effect of the data transformation measures (what the Commission calls “technical measures”) and the controls on the recipient’s environment (which the Commission calls “contractual measures” but which cover technical, organisational and legal measures). We examine whether these combined measures could be deemed sufficient to reduce the risk of re-identification to an insignificant level in reality in the light of the European Data Protection Board (EDPB) standard in its latest version of its anonymisation guidance [4].

The contributions of this study are as follows: (a) a critical analysis of the anonymisation measures specified by the Commission against the EDPB guidance; (b) an empirical evaluation of the residual data vulnerability through a motivated intruder test on the Search Dataset; (c) a quantitative modelling framework for assessing the re-identification risk using results from the motivated intruder test and the controls of the recipient’s environment with a view to assess the Commission’s anonymisation claim through the lenses of a contextual approach to anonymisation. These contributions address the immediate question of whether the specified measures can support the that the Search Dataset is anonymised for the population of end users. Beyond this Alphabet-specific case study, they offer broader lessons about re-identification risks in search query data and motivated intruder tests for evaluating those risks.

# 2. Background

## 2.1. Principles of Risk-Based Anonymisation

Analysing anonymisation measures requires considering both the information retained in the data and the circumstances in which it will be accessed and used. Anonymisation is best understood as a risk management process involving iterative assessment, mitigation and ongoing review. Its objective is to establish that residual re-identification risks meet the applicable legal standard. In practice, where data are being anonymised for further use, preserving an acceptable level of utility is also essential.

This understanding is reflected in the literature [5–8], international standards (e.g., ISO/IEC 27559:2022) [9] and regulatory guidance across jurisdictions [4,10–12]. Although jurisdictions differ in their terminology and requirements for considering information anonymised [13], their identifiability tests commonly incorporated contextual qualifications, expressed through formulations such as "reasonably foreseeable", "serious possibility", "reasonably likely to be used" or "reasonably be expected" [13]. These qualifications direct attention to identification possibilities that are realistic in the circumstances, rather than by considering every theoretical possibility. A re-identification risk assessment therefore determines whether residual identifiability meets the applicable legal standard, without necessarily requiring a zero risk of re-identification.

In practice, this requires assessing the characteristics of the data alongside assumptions about potential adversaries, their capabilities and knowledge, access conditions, and the effectiveness of contractual, organisational and technical controls [8]. This contextual understanding is conceptually consistent with the Five Safes [14], a framework for governing safe data access and use through joint consideration of projects, people, data, settings and outputs.

In Europe, the Article 29 Working Party's 2014 Opinion acknowledged the possibility of a thorough evaluation of re-identification risks and highlights contextual factors, combinations of techniques and regular reassessment of residual risks [15]. The EDPB Guidelines 02/2026 further develop this approach by explaining how anonymity should be assessed from the perspective of the relevant entities, taking account of the means reasonably likely to be used to identify individuals [4]. While the guidelines recognise that technical, organisational and contractual measures can restrict access, they also emphasize that they do not themselves establish anonymity. Their contribution must be evaluated in relation to the identification capabilities of the relevant entities as well as the transformations that are performed on the data itself.

European guidance has articulated three criteria for assessing anonymity since the Article 29 Working Party 2014 Opinion: singling out, linkability and inference [15]. The EDPB's 02/2026 guidelines refine these as No Record Isolation, No Linkage and No Inference [4]. They examine whether individual records can be isolated, information concerning the same individual can be linked across datasets, or specific and meaningful information about an identified or identifiable individual can be inferred. Satisfying all three supports a finding of anonymity; failing a criterion requires further analysis rather than automatically establishing identifiability. Similar criteria extend beyond Europe: Quebec's anonymisation regulation expressly requires consideration of individualisation, correlation and inference [16].

From a technical perspective, the privacy risks that anonymisation seeks to mitigate are commonly described through three disclosure outcomes: identity, attribute and membership disclosure (see also relevant terminology as listed in **Table 1**). Identity disclosure concerns correctly associating an individual with their record; attribute disclosure concerns learning characteristics of an individual from the dataset's attributes; and membership disclosure concerns learning that an individual contributed to a dataset. For inferential disclosures, the baseline against which information gain is assessed is important. An adversary may already infer an attribute from background knowledge or population-level patterns [17–19] or, in case of membership disclosure, have prior knowledge about an individual's inclusion [20]. The inference assessment therefore concerns the additional information that the released data reveals about that individual beyond this baseline.

The relationship between regulatory anonymisation criteria and technical disclosure concepts is not straightforward and has been debated in the literature [17,21–23]. In general, singling out and linkability can enable identity disclosure, whereas attribute and membership disclosure describe information that may be inferred about an individual, potentially without identifying their specific record in the dataset [18,20,21,24,25]. However, isolating a unique record, recognising the same individual across interactions and establishing a person-to-record association are distinct capabilities. The EDPB Guidelines 02/2026 distinguish record isolation from singling out an individual: a record may be unique within a dataset without enabling a relevant entity to identify the person concerned (see Example 24 in [4]). Conversely, recognising an individual across interactions may constitute identification even without knowing their name (see Example 6 in [4]). This distinction is also highlighted in the academic literature, which differentiates singling out from establishing a person-to-record association [5,26].

## 2.2. Disclosure Risk Assessment and Mitigation

For a specified disclosure scenario, re-identification risk can be decomposed into the probability that an attack occurs and the conditional probability of successful disclosure given that attack [9,27]:

$$R = P(A) \times P(D|A) \tag{1}$$

where $P(A)$ is the probability of an attack occurring and $P(D|A)$ the disclosure vulnerability of the data given that that attack. Here, “attack” is used broadly for any event where disclosure may occur, including deliberate identification attempts, inadvertent recognition during legitimate use (i.e., spontaneous recognition) and attacks following a breach [9].

This decomposition identifies two levers for mitigation. Data transformations reduce vulnerability; and organisational, contractual and technical safeguards reduce the probability of attack by, for example, restricting access to the data or to additional information needed for identification, through the screening and training of individuals. Their contribution nevertheless depends on the release conditions and their practical effectiveness. As the EDPB emphasizes, safeguards do not themselves establish anonymity [4]. [1]

Data transformations include removing direct identifiers (i.e., pseudonymisation), generalising, suppressing or perturbing quasi-identifying information [5], and synthetic data generation [18]. The transformation measures applied to the Search Dataset examined in this study are described in **Section 2.4.**

Assessing the effectiveness of the transformations and the safeguards requires evaluating residual disclosure risk. Privacy is understood here primarily as informational privacy [5] or better confidentiality, as one goal to meet within the EU data protection framework. The assessment therefore concerns successful disclosure, rather than downstream harms, which may arise from either successful disclosure or incorrect claims of re-identification or attribution.

Disclosure risk can be assessed through qualitative approaches, quantitative modelling and empirical motivated intruder tests [9,10]. Qualitative approaches typically rely on heuristics, although such fixed rules may provide insufficient protection in some datasets and unnecessarily reduce utility for a given context in others. Quantitative modelling estimates vulnerability and risk under explicit assumptions about the data, adversaries and disclosure scenarios, allowing comparison with a justified threshold. A motivated intruder test complements modelling by empirically investigating whether plausible disclosure attempts succeed in practice. It is the focus of this study and is described in **Section 4.**

Quantitative identity-disclosure assessment commonly uses equivalence classes: groups sharing quasi-identifying information assumed to be available to an adversary [9]. Larger groups allow individuals to hide among more matching candidates. Individual vulnerability estimates can be summarised through maximum or average measures. Maximum vulnerability captures the most vulnerable individual under the model and is particularly relevant to public releases. Under a simple matching model, k-anonymity bounds this vulnerability by requiring a minimum group size [28]. Average vulnerability represents expected vulnerability for a randomly selected individual. ISO/IEC 27559:2022 supports vulnerability assessment in non-public release settings where effective controls and safeguards support the assumed disclosure scenarios [9]. These methodological recommendations do not themselves establish legal anonymity but provide compelling precedent frameworks for assessing disclosure risks.

[1] For public release scenarios, enforceable controls over recipients are generally unavailable, and therefore risk mitigation must rely primarily on data transformations. However, in the current context, this is not a public release and hence the role of these safeguards is important.

**Table 1:** Terminology used in this paper.

| Term | Meaning used in this paper |
|---|---|
| **Privacy Concepts** | |
| Adversary | An adversary (or attacker, intruder) is an individual whose access to data may lead to re-identification, deliberately or inadvertently. Typically modelled as the anticipated recipient. |
| Anonymisation | A process of assessing and mitigating identifiability to produce data meeting the applicable legal standard of anonymity in the relevant context. |
| Attribute disclosure | Inferring information about an individual's attributes without necessarily identifying their specific record. |
| Direct identifier | Information that directly identifies an individual in the relevant context, such as a name, email address or identification number. |
| Disclosure vulnerability | The conditional probability that a specified disclosure succeeds given the relevant event, under explicit assumptions about the data and the adversary's knowledge and capabilities. |
| Disclosure risk | The overall probability of a specified disclosure outcome through a defined scenario, combining event probability and conditional vulnerability. |
| Equivalence class | A group sharing the same QI values and therefore indistinguishable under the specified thread model. |
| Identity disclosure | Correctly associating an individual with their corresponding record(s) in the data. |
| Inferential disclosure | Learning information about an individual beyond an appropriate prior-information baseline, with or without identity disclosure. Encompassing attribute and membership disclosure. |
| Membership disclosure | Learning that an individual is represented in a dataset. |
| Pseudonymization | Processing of direct identifiers that prevents direct attribution to a specific individual. |
| Quasi-identifier (QI) | Information that can contribute to re-identification in combination with other information and is plausibly available to an adversary. |
| Re-identification | Successful association of an individual with their record(s). In general, an individual's record is reidentified rather than the individual—an individual exists independently of the data release and is already "identified" in the real world. |

## 2.3. Anonymisation and Risk Assessment in Search Dataset

Applying these approaches to the Search Dataset presents additional challenges. Query logs typically combine structured metadata, such as timestamps and clicked URLs, with unstructured query text that may contain names, locations, affiliations, health conditions and other personal details. Such information does not necessarily concern the person issuing the query: searches may relate to relatives, public figures or hypothetical situations.

Nevertheless, combinations of queries and metadata can provide distinctive clues supporting identity disclosure. This became publicly apparent in 2006, when AOL released approximately 20 million search queries from around 650,000 users with user identifiers replaced by pseudonyms. Journalists subsequently identified an individual from the content of her query history [29].

Early research demonstrated several mechanisms underlying these vulnerabilities. Jones et al. [30] showed that query histories could reveal demographic characteristics and narrow candidate identities, while combinations of interests could help locate an individual's history. Kumar et al. [31] demonstrated that statistical analysis could partially recover query content protected through token-based hashing. These findings illustrate why removing direct identifiers is insufficient to protect against identity disclosure. Proposed more comprehensive anonymization approaches include deleting identifiers, scrubbing personal information from query text, suppressing infrequent queries and limiting linkage across a user's searches [32]. More formal approaches include user-level k-anonymity that groups similar user histories [33,34] and releasing aggregate query and click statistics under variants of differential privacy [35,36].

Assessing residual disclosure risk in query logs requires considering query text, structured metadata and longitudinal patterns jointly. The disclosure may concern an individual's identity, attributes or membership in a dataset – identity, attribute or membership disclosure, respectively. Existing studies examine inference [30,37], linkage between transformed and original histories [34] and to cluster otherwise disconnected user records together [38]. Although these outcomes reveal potential vulnerabilities, they do not necessarily establish that disclosure has occurred or quantify its likelihood under a specified disclosure scenario. Empirical evidence quantifying residual disclosure risk in transformed query logs under realistic adversary conditions remains limited [39].

## 2.4. Alphabet Search Data and its Anonymisation

Under Article 6(11) DMA, Alphabet is required to share their "ranking, query, click and view data in relation to free and paid search" in anonymised form with third party search engines [3]. The scope specified in the Commission's July 2026 decision covers queries entered into Google Search, the results displayed, their ranking and users' interactions with them. The underlying data also include query modifications and metadata such as location, device type and account identifiers. These fields are subject to the suppression and transformation measures described below. The data cover searches by both signed-in and signed-out users [3].

Alphabet has offered access to anonymised search data through its European Search Dataset Licensing Program since March 2024. The Commission's July 2026 decision specifies changes to that existing arrangement, including the scope of the data and the anonymisation measures required before sharing [3] (recitals 5-9). These measures give effect to the second sentence of Article 6(11) DMA which provides that "[a]ny such query, click and view data that constitutes personal data shall be anonymised." The technical measures alter the data in four steps:

1) Attribute suppression removes end-user, IP and device identifiers and the query timestamp, retaining only the day.
2) Query suppression removes any query containing an entity below an allow-list threshold, which admits an entity only if more than 50 signed-in end users issued queries containing it over the preceding 13 months of European Economic Area (EEA) data, and removes queries longer than the 95th percentile of query length for their language.
3) Metadata generalisation reports location at NUTS 3 level, requires at least 1,000 signed-in users per language-location-device group per day, and bins interaction durations by within-group percentiles.
4) Mini-sessionisation groups at most three refinement-related queries under a random identifier, so that the released data contain short fragments rather than per-user histories. Contractual measures, including segregation, prohibitions on re-identification, linking and augmentation, and logging and audit requirements, address the risk that remains [3].

Between the April provisional measures and the July decision, the Commission raised the level of confidentiality assured to end users in two main ways: by shrinking what can be learned from each record and by tightening the recipient's environment. The metadata threshold was moved from 50 to 1,000 users, location became NUTS 3 with a country-level fallback, and further attributes were stripped (access point, input format, ad URLs, advanced filters, account-linked results, AI Mode follow-ups). Interaction timings were coarsened and mini-sessions capped at three queries. A mandatory segregated environment, a further specified ban on linking, augmentation and re-identification and earlier audits (first Level 2 report after six months, plus ad hoc reports) shift more of the residual risk onto verifiable controls. The core design of the Commission's measures remain unchanged, however: query text is still filtered by an entity threshold of 50 signed-in users and a length cap.

Notably, eligibility moved in the same direction as confidentiality. The April draft relied on the definition of an online search engine (OSE) in the Platform-to-Business Regulation [40], whereas the July decision adds explicit screening: third-country control, a genuine-economic-activity test, a transfer impact assessment if relevant, a broad equivalence test, and a public-security exemption route (Annex paragraphs 3-9). This narrows the pool of recipients, which supports confidentiality. It also increases Alphabet's gatekeeping role in this regard (limited by the FRAND, i.e., fair, reasonable and non-discriminatory, requirement and Commission oversight) and may exclude very small entrants, creating a tension with the contestability goal.

The available evidence on the robustness of the Commission's anonymisation approach includes empirical testing by both the Commission and Alphabet, although its interpretation remains contested. The Commission's testing examined names and addresses in query text, used large language model (LLM)-assisted clustering to evaluate re-sessionisation, and manually assessed selected reconstructed sessions for re-identification potential [3] (recitals 748–758). For example, the Commission reported that clustering accuracy for moderately similar queries was as low as 10% in groups of 1,000 queries and approached random guessing for dissimilar queries [3] (recitals 757–758). The Commission's testing quantified aspects of within-dataset linkability and examined whether potentially identifying information could be attributed to the person issuing a query. Alphabet conducted broader adversarial testing of the April preliminary measures on one week of Irish data, involving 10 to 15 engineers and external experts over at least two months, and reported identifying several individuals in under two hours [3] (recital 721). Its tests included attempts at identification using query content, metadata and auxiliary information, as well as re-sessionisation. The Commission challenged the representativeness and realism of these tests and revised several measures in response to the vulnerabilities identified [3] (recitals 713, 722–746). These findings provide evidence about specific disclosure mechanisms, but neither establishes a systematic evaluation of automated identity disclosure at scale or reports a demonstrated attack starting from a known individual.

# 3. Critical Assessment of the Commission's Approach to Anonymisation

Under the Commission's implementing decision, the technical transformations must reduce re-identification risk to a *residual* level. Complementary contractual measures are intended to reduce the remaining risk to an *insignificant* level. The Commission's risk assessment follows the Expert Report on Anonymisation, which treats a risk as beyond the residual level if any one of three criteria is met [3] (recitals 618–619):

a) an adversary can choose a specific individual and, with significant probability, retrieve some of their search records with high certainty;

b) an adversary can choose a random search record and, with significant probability, re-identify the end user with high certainty; or

c) an adversary can re-identify a large number of end users with high certainty using reasonably simple methods.

The effort required in terms of time, cost, manpower and technology, and the information the adversary is expected to gain, form part of the assessment [3] (Recital 619). While these criteria aim to operationalize the concepts used in EU anonymisation guidance, they raise several questions that should be set out before the analysis begins, so that its scope and limits are clear.

## 3.1. Analysis of the Commission's Requirements in the Light of EU Anonymisation Guidance

The anonymisation standard under the GDPR has long resisted clear definition. Tensions within early guidance on anonymisation, particularly Opinion 05/2014 [15], show that even data protection authorities have struggled to formulate clear and consensual guidance on the anonymisation standard. The test derived from the CJEU's case law remains much debated, not least because the Court has, understandably, decided cases on narrow grounds [41–44]. This unsettled background gives particular weight to Article 6(11) DMA specification proceedings, which require the Commission, once proceedings are opened, to define ex ante the measures that the gatekeeper must implement. Now that the EDPB Guidelines 02/2026 on anonymisation have been released,[4] the Commission's approach should be assessed against them, to test the Commission's claims more closely.

### 3.1.1. Protecting only end-users

The Commission's anonymisation measures are part of a broader strategy: making search data sharing workable for rival search engines, so that the provision delivers contestability, while aligning the anonymisation standard with GDPR requirements. The Commission's implementing decision of 16 July 2026 sets out more explicitly than its preliminary findings of 16 April what the Commission is really trying to achieve for Article 6(11) DMA [45]. Although some formulations in the Commission's decision may remain imprecise [3] (see e.g., recital 121(c)), in substance, however, the Commission is specifying the measures Alphabet must implement to anonymise end-user data, that is, to protect the confidentiality of end users only.

In recital 551, the Commission distinguishes end users from other individuals represented in the data, and states that “Article 6(11) of Regulation (EU) 2022/1925 read in the light of recital 61 of that Regulation provides that the anonymisation requirement applies to the personal data of the end user generating the data and not to the personal data of other individuals that might be identifiable via ranking, query, click and view data (e.g., personal data mentioned in query and ranking data).” Each of criteria (a) to (c) refers to retrieving the records of, or re-identifying, an end user, and the test of residual risk is applied to that population alone. The Commission concludes that Online Search Engines (OSEs) are "responsible and liable for the processing of personal data of individuals other than end users forming part of the Search Dataset under Regulation (EU) 2016/679" [3](recital 807). OSEs are independent controllers [3] (recital 809 and annex paragraph (45).

The Commission’s anonymisation claim is therefore only partial. As a result, although the anonymisation measures utilise personal data detectors in general, that is irrespective of the population of individuals represented within the data, the Commission’s entity-based threshold is defined to remove “secrets and likely unique identifiers such as passwords, private links and social security numbers,” [3] (recital 651) but not to remove all personal data. With this said, the Commission’s approach confirms that it is difficult in practice to distinguish between end-users and other individuals when defining the anonymisation criteria and making claims per data subject type. A search about a health condition, for example, may concern the end user, their child, a colleague or simply a topic they encountered while reading. The query alone therefore establishes neither that the condition relates to the end user nor that any particular person has it. This ambiguity complicates both the application of anonymisation criteria that distinguish end users from other individuals and the interpretation of apparent attribute disclosure.

### 3.1.2. Protecting all end-users

A crucial question is therefore what type of guarantee the Commission’s anonymisation techniques ultimately provide, and whether they can show, as the EDPB appears to require for an anonymisation claim to succeed, that the likelihood of re-identification is insignificant for all end users, even if no equal treatment is required [4] (paragraph 36). Read against the background of the EDPB Guidelines 02/2026, the Commission therefore would be suggesting that an insignificant risk is only achieved for all end users through a combination of technical and contractual measures [3] (recital 789). That approach does not appear, in principle, at odds with the EDPB’s guidance.

Where the Commission is not clear is about the exact effect of the technical measures. On one hand it states that “these measures alter the Search Data and play a prominent role to mitigate the risk of re-identification of end users without substantially degrading the data quality and usefulness” [3] (recital 593(a)). On the other it states that they “are indispensable and play a prominent role in reducing the risk of re-identification to an insignificant level” [3] (recital 618) and more precisely that they “reduce the overall re-identification risk to a residual level without substantially degrading the data quality and usefulness and the contractual measures further mitigate that residual risk to an insignificant level [3] (recital 618). It then adds that the Expert Report on Anonymisation that informed its work considered that the technical measures reduced the risk beyond residual level [3] (recital 619). Taken together, these statements suggest that the Commission considers all three re-identification risks to remain, at most, at a residual level following the application of the technical measures, with the contractual measures subsequently reducing them to an insignificant level. This approach to managing risk is consistent with contemporary guidelines and standards [46,47].

Moreover, despite its broad assertions, the Commission’s reasoning appears to rely on an average-risk approach. Reducing risk to a residual level through technical measures, and then to an insignificant level through additional controls while relying upon an average risk metric is consistent with established best practices for contextual approaches to anonymisation. This, however, raises questions as to how the Commission's position can be reconciled with the EDPB Guidelines 02/2026. First, an average-risk assessment may leave room for outliers: some individuals whose data are included in the Search Dataset may remain exposed to a higher risk of re-identification than the average. Second, the Commission does not appear to identify a specific metric or threshold demonstrating that the technical measures reduce each of the three re-identification risks to a residual level for all end users, before the contractual measures bring those risks down to an insignificant level for all end users. The distinction is important: demonstrating that risk is insignificant on average is not the same as demonstrating that it is insignificant for every end user. This also highlights the important complementary tole that the contractual measures play in ensuring an insignificant level of risk.

### 3.1.3. No Record Isolation and No Singling Out

Measured against the EDPB Guidelines 02/2026,[2] the Commission takes a contextual approach, which it claims assesses singling out, linkability and inference from the perspective of OSEs. It rejected calls for formal approaches that apply k-anonymity to metadata and query text together in order to preserve utility and pursue the DMA's contestability objective.

Even though the Commission's vocabulary remains that of Opinion 05/2014, i.e., singling out, inference and linkability, [3] (recital 621) the Commission holds that isolating some or all records relating to a specific end user, including by linking queries through re-sessionization, does not by itself constitute successful re-identification, because the isolated records must also be reliably linked to a particular individual [3] (recital 622(c)).

The EDPB Guidelines 02/2026 defines the No Record Isolation criterion as a unique record in the data [4] (paragraph 55). In the case of the Search Data, it is a reasonable argument to make that this represents the whole population, and therefore a unique record in the data is also unique in the (whole) population and could enable re-identification if linkable additional information makes it possible to identify the individual. Paragraphs 98-99 of the EDPB Guidelines 02/2026, however, could be interpreted in at least two ways [4] (see also [48,49]): either singling out is established when a record is unique within the (whole) population or when a record is unique within the (whole) population and when linkable additional information is reasonably accessible to the relevant entities. The fact that the EDPB expressly states that "if the data violates the No Record Isolation criterion, it should be checked if the isolated records allow for singling out," is not enough to definitely eliminate the ambiguity [4] (paragraph 98).

In practice, linking a record to an identity (i.e., linkable additional information is reasonably accessible to relevant entities) is an important condition for re-identification to occur. As will be seen in the results of our analysis, suspected matches are often not correct or not verifiable. Therefore, uniqueness alone would not be sufficient to establish that re-identification has occurred.

It should be acknowledged that linking a unique record to an identity is a significant step with many points of failure. Uniqueness therefore cannot always be treated as sufficient. Given the characteristics of the use case at hand, there is a reasonable argument that re-identification is established only when a unique row within a population is actually linked to an identity. This view matches the two-step logic of the EDPB Guidelines, under which an isolated record counts as singling out only if it can be matched to an individual's attributes (paragraphs 98 to 99). It also keeps the test aligned with recital 26 GDPR, which asks whether identification is reasonably likely, not whether it is theoretically conceivable. Treating uniqueness as a necessary indicator, to be followed by an assessment of whether the link to an identity can realistically be made and whether such a link matters for the given context (e.g., which is arguably not the case in situation of profiling), avoids two errors: classifying every unique record as re-identified for all contexts, and ignoring uniqueness altogether.

[2] Some of the feedback provided during the consultation period have identified inconsistencies and made suggestions for changes, and it is therefore possible that a new version of the guidelines would be different in features material for the current discussion [48,49].

### 3.1.4. No Singling Out and No Inference

Not only are the EDPB Guidelines 02/2026 ambiguous on how singling out is defined but also on how it operates in relation to other risks. They suggest that No Singling out can only offset No Record Isolation, yet they define specific and meaningful inferences as inferences about identified or identifiable individuals [4] (paragraph 67). In principle, therefore, No Inference remains in scope under a contextual approach. Attribute disclosure is of particular importance here. Membership disclosure, establishing that an individual's data was included in the source dataset, is another inference that in this context, however, does not provide meaningful information because it is reasonable to assume that the Search Dataset includes the whole population.

A crucial question is therefore whether the Search Dataset permits specific and meaningful (attribute) inferences to be drawn about individuals, and more specifically what it really implies to state that an inference is only specific and meaningful when the individual, i.e., the end user in the case of the Search Dataset, is identified or identifiable, as the EDPB writes.

In principle, assessing an inference risk requires distinguishing information gained through the release from information already inferable from population patterns or other reasonably available resources. In the context of the end-users in the Search Dataset, the Commission's anonymisation measures seem to imply that identity disclosure is a pre-requisite to attribute disclosure.

The bucketing mechanisms and entity thresholds may constrain the possibility of attribute disclosure, but group size and entity frequency alone do not establish protection. Nor should unusual queries or rare combinations of characteristics automatically be treated as the most vulnerable: their distinctiveness may facilitate identity disclosure, while attribute disclosure depends on the sensitive information and its correlations with the available quasi-identifying information. Conversely, common characteristics may support substantial attribute inference.

Testing these risks requires appropriate baselines and reliable information against which inferred attributes can be validated. Query content alone provides limited ground truth because it need not describe the person issuing the query. Tests focused on identity disclosure therefore provide only part of the evidence needed to assess inference risks. The assessment should also consider whether the Search Data enables information to be inferred about identified or identifiable individuals beyond what could be inferred from population-level information, even where no specific search record has been linked to them.

The Commission's claims regarding the risks of inference are particularly difficult to test empirically. User account data might help validate inferred attributes, but would not by itself establish whether those inferences depend on the individual's inclusion in the Search Dataset rather than on population-level patterns. Assessing this dependence through comparisons with and without the individual's data requires access to the underlying data and anonymisation process, beyond what is available to a motivated intruder. Moreover, the criteria (a) to (c) serving as the operational test of whether residual re-identification risks persist, are not specifically built for testing inference risks.

One way to justify the Commission's position may be to argue, referring to the EDPB's position, that "[a]n inference is specific if the inferred information relates to a single identified or identifiable individual" [4] (paragraph 69), and that the Commission is addressing inferences through its focus upon singling out and the criteria (a) to (c). This suggests however that singling out is understood to be met when a record is unique within the (whole) population and linkable additional information is reasonably accessible to the relevant entities.

### 3.1.5. Mixed data

Importantly, the Commission leaves space for an independent third-party assessment, which can be grounded in the EDPB Guidelines, to continue treating the Search Dataset as personal data as a whole. The EDPB Guidelines 02/2026 confirmed the importance of sticking to a global approach when making findings of anonymisation and distinguish between record-level and database-level anonymisation claims in line with the CJEU's reasoning in Meta v Bundeskartellamt [50] (paragraph 89). The EDPB confirms that "[s]ome anonymisation techniques might only make the likelihood of identification insignificant for some individuals in a dataset. In such cases, it is important to distinguish between the anonymity of particular records and the anonymity of the dataset per se"[4] (paragraph 36).

As mentioned above, the Commission does not make a database-level anonymization claim, which suggests that the Search Dataset "contains a mix of personal and anonymous data," so that "the entire dataset should be treated as containing personal data"[4] (paragraph 36). In a letter addressed to the Commission on the 5 May 2026 the EDPB acknowledges the limited anonymisation goal set by the DMA: "the EDPB underlines that anonymisation in the context of Article 6(11) DMA, and therefore also in the context of the proposed measures, concern only anonymisation of the personal data of the end user generating the data" [51]. On this basis, the EDPB concludes that the GDPR remains fully applicable to the processing of personal data relating to individuals, including end users, who are searched for through a query. It is precisely because the Search Dataset does not allow for a distinction to be drawn between end users and other individuals represented in the dataset that it should be regarded as a mixed dataset. This may explain why the Commission accepts that the contractual measures will ultimately absorb differential treatment without introducing additional criteria. In any case, the EDPB remains elusive on whether a formal maximum limit on re-identification risk is needed in all cases.

### 3.1.6. Outstanding questions

Three questions must therefore be distinguished: (1) what the anonymisation techniques actually achieve in terms of privacy, considering No Record Isolation, No Linkage, No Inference and No Singling Out which does a lot of work under the EDPB Guidelines 02/2026 ; (2) what the addition of contractual controls makes it possible to achieve in terms of privacy, considering No Record Isolation, No Linkage, No Inference and No Singling Out; and (3) what the legal status of the Search Dataset is in the hands of the OSEs.

In this paper questions (1) and (2) are assessed against existing standards for a contextual approach through a motivated intruder test, noting where relevant the limits of the analysis imposed by the conditions under which the data are accessed. The EDPB Guidelines 02/2026 seem to acknowledge the relevance of a motivated intruder test, while cautioning against assuming too readily that an intruder would lack motivation [4] (paragraph 31). Question (3) is touched upon within the discussion.

## 3.2. Analysis of the Context Controls Set Upon the Environment of the Search Dataset

The contractual controls the Commission introduced to complement the technical measures are central to the anonymisation claim. As noted above, they are what allow the Commission to conclude that the risk of re-identification is ultimately insignificant. Controls of this kind are standard in contextual approaches to anonymisation and appear explicitly in regulatory guidance in several jurisdictions, for example Ontario's De-identification Guidelines for Structured Data, revised in 2025 [10]. Because this paper focuses upon assessing the strength of the Commission's contextual approach, it must examine these controls as well.

While the EDPB appears cautious about the role of contractual controls in substantiating contextual anonymisation claims, it acknowledges that security measures applied to the data environment can support a conclusion that unauthorised actors do not have access to re-identification means reasonably likely to be used [4] (paragraph 89). This recognises the importance of technical contextual controls, which, without contractual obligations and a higher level security duty, would not be binding [49]. .

The Commission's controls are grouped into three main categories: controls for ensuring anonymisation from OSEs' perspective, controls for ensuring integrity and confidentiality to ensure anonymisation from all other persons' perspective, and an independent verification mechanism to ensure anonymisation over time. In line with the EDPB's position [4] (paragraph 89), the Commission considers that the integrity and confidentiality measures mean that access to the Search Dataset should not be deemed a means reasonably likely to be used by unauthorised users.

The Commission seems to go further than the EDPB, however. In its view, multiple safeguards support the conclusion that the Search Dataset should not be regarded as a means reasonably likely to be used by OSEs to identify individuals. These safeguards comprise: OSEs' sole control over their segregated processing environments, which must be logically and technically segregated from the rest of their processing capabilities, the prohibition against linking, re-identification and augmentation, the restriction to permitted purposes of use to ensure that the Search Dataset is only processed for the purposes of optimising and improving OSE services, the testing safeguard for ML models to address memorisation risks, the prohibition of downward sharing and disclosure, and the governance and traceability measures.

Compared with the April provisional measures, the July decision makes the contractual safeguards easier to test and enforce. First, recipients must process the data in a logically and technically segregated environment, where it cannot be queried jointly with, or copied, merged or synchronised into, any other dataset, and where subcontractors must also work (Annex paragraph 49). The April provisional measures only mandated organisational measures at the infrastructure level to separate the Search Dataset from auxiliary advertising and analytics datasets. Second, the ban on re-identification now covers specific techniques, such as semantic similarity and behavioural fingerprinting, used to rebuild a user's history beyond the three-query mini-sessions (Annex paragraph 50(b)). Third, while the April provisional measures did not mention confidentiality obligations or training, however, the July decision includes an express reference to them (Annex paragraph 85(d)). Fourth, the audit comes sooner and can be triggered on demand. The first operating-effectiveness (Level 2) report is due six months after access, based on a three-month sample, instead of 15 months (Annex paragraph 139(a)). The Commission can also require an ad hoc report at any time (Annex paragraph 140). Significantly, the threshold for expedited termination of access is easier to meet with the July decision. When the April provisional measures required an "urgent risk of serious and irreparable damage to anonymisation" for expedited termination, the July decision requires "substantiated evidence demonstrating that the integrity and confidentiality of the Search Dataset has been compromised" (Annex paragraph 164(a)).

Compared with existing guidance on contextual approaches, the Commission's controls appear fairly comprehensive. To test this, we mapped each control to the checklist of mitigating controls in Appendix A of the Ontario De-Identification Guidelines for Structured Data (IPC Guidelines). The complete mapping is available in **Appendix A**. Controls were matched by functional equivalence rather than wording (see the Annex). We coded 24 distinct contractual and environmental controls (C1–C24) by reading the Annex of the Commission's implementing decision paragraph by paragraph and cross-referencing each control to the recitals that explain it. Controls on the data itself (the technical transformations) were excluded.

Each control corresponds to a single obligation or tightly linked set of obligations. We then mapped each control to the 31 items of the checklist in Appendix A of the IPC Guidelines, reading each item's description rather than its heading. We also refer to Appendix B and C where relevant. Of the 24 controls, six (C6, C8,

C10, C13, C14, C16) map closely to Appendix A and one (C24) to a clause in Appendix C. Eleven (C4, C7, C9, C11, C12, C15, C17, C18, C21, C22, C23) map partially but still relatively well and more often than not this is because the Commission's controls are more specific. Three (C1, C2 and C3) map weakly but this is because the Commission's controls are more restrictive. Three controls have no counterpart in Appendix A: the testing safeguard for machine learning models (C5) and the obligations on change of control and insolvency (C19 and C20). The first responds to a risk specific to model training on the Search Dataset, the last two aim to keep the contractual measures enforceable when the recipient's circumstances change.

Overall, although the terminology differs and the wording is in places less specific (see, for example, the provisions on physical security) and in others more specific (see, for example onward sharing and restricted permitted purposes of use), the Commission's controls cover much of the same ground as the controls listed in Appendix A of the IPC Guidelines. Importantly, the Commission's controls are designed for a single, specific use case: the reuse of data to optimise and improve OSE services. In that respect they appear more restrictive than the IPC guidance, which is more generic. Because the purpose is already fixed, several generic governance controls that govern the choice of reuse purpose become largely redundant. In effect, the resulting package reads like the mitigating measures one would expect to find in a data protection impact assessment although more details would still be needed.

Two controls show most clearly how the Commission's approach departs from generic guidance: permitted purposes (C4) and the prohibition on onward sharing (C7). Appendix A of the IPC Guidelines has no purpose restriction item. It addresses what data may be used for through governance (an independent oversight or stewardship committee that defines the scope of processing) and through transparency about analysis objectives, while Appendix C adds that data sharing agreements commonly define permitted and proscribed purposes, negotiated between the parties. Under the Commission's decision, the purpose is neither negotiated nor set by a committee: Article 6(11) DMA fixes it as the optimisation and improvement of the recipient's own search services. The decision therefore dispenses with controls on the choice of purpose and concentrates on policing it, through a contractual purpose clause, documentation of the systems and models derived from the data (C8), pre-deployment testing of models for memorisation (C5) and independent assurance (C17–C18). The same logic seems to explain the treatment of onward sharing. The IPC Guidelines allows recipients to share with third parties if the data controller gives prior approval and the same controls travel with the data (Appendix A and Appendix C), whereas the Commission prohibits onward sharing outright, except to processors and sub-processors that are bound by the same restrictions and open to audit (C24). A prior-approval route would give Alphabet discretion over competitors' use of the data, which sits uneasily with the decision's limit on extending control over recipients' independent processing and with the contestability objective.

On data segmentation specifically, the IPC Guidelines accept that credible, documented and enforced separations can be built between identified and de-identified data (separate teams, access limits, technical or physical silos) to support reuse within an organisation (Section 4.5 and Appendix B), while the ICO, which we used as a reference point for the motivated intruder test, by contrast, appears to reason at the entity level, writing that "if you can still identify people using additional information you hold separately, the data is not anonymised" [52].

The relationship between Alphabet and data beneficiaries is likely more adversarial than the drafters of the IPC Guidelines envisage. This may explain why in endorsing data segmentation through the requirement of a segregated processing environment coupled with a prohibition on linking, the Commission stresses that the contractual measures stem from the regulator's implementing decision itself, although, ultimately, as mentioned above, the Commission does not claim that at the dataset level the measures would amount to a change of status, meaning that the Search Dataset would then be considered as a whole anonymised.

It is thus important to stress that OSEs remain subject to data protection law as independent controllers. As the Commission states it, "the fact that the Search Dataset will inevitably contain personal data relating to other individuals than the end-users necessarily entails the continued application of Regulation (EU) 2016/679 to the sharing of the Search Dataset and to the subsequent processing operations carried out by third-party OSEs" [3] (recital 800). This implies that OSEs remain subject to the GDPR for the Search Dataset as a whole, including the whole set of data protection principles and international transfer restrictions.

Notably, Alphabet is itself subject to a transparency obligation. Through its website, it must inform data subjects of the OSEs with which the Search Dataset is shared, of the existence of level 1 and level 2 assurance reports, and, where relevant, of any termination of an OSE's access to the Search Dataset [3] (recital 877).

In addition, the Commission acknowledges that OSEs will have to establish their own legal basis under Article 6 GDPR and points to legitimate interest as a potential candidate [3] (recital 809). This raises at least two related questions. First, a question about operationalization of safeguards, such as data subject rights and in particular the right to object, which is arguably inexistent prior to the sharing of the Search Dataset. Second, a question about appropriateness of the legal basis assuming special categories of personal data are present in the Search Dataset. Yet, such a presence cannot be excluded. There might be space, however, for two types of arguments: 1) an argument made on the basis of Article 9(2)(g) GDPR, even if the Commission only mentions Article 9(2)(g) GDPR in relation to the sharing of the Search Dataset by Alphabet when it states that "[t]hat contestability objective constitutes a substantial public interest recognised by Regulation (EU) 2022/1925 within the meaning of Article 9(2)(g) of Regulation (EU) 2016/679" [3] (recital 797); or perhaps less convincingly 2) an argument following the line of reasoning of the French CNIL applied in the context of the training of Artificial Intelligence models in an attempt to save incidental and residual processing of special categories of personal data [53].

Lawfulness is not the only data protection principle that will require further work on the part of OSEs. Purpose limitation calls for more specificity than the description of the permitted purposes provides, and the same is true of data minimisation.

Ultimately as regards redress, the Commission acknowledges that if a data breach occurs, end user data lose its partially-anonymised status. From this consideration, it concludes that the risks created to data subjects is not beyond legal redress [3] (Recital 884).

## 3.3. Summary of Analysis of Commission Measures

Alphabet is bound by the implementing decision and must, in turn, impose a set of technical, organisational and legal measures on OSEs by contract. Compared with state-of-the-art guidance on contextual controls, particularly the IPC Guidelines, the Commission's measures appear fairly comprehensive. They circumscribe reuse of the Search Dataset through permitted purposes of uses and a segregated processing environment that must generate its audit trail, and they prohibit attempts to re-identify end users, to link the data with auxiliary datasets or to augment the Search Dataset so that it reverses, weakens, or circumvents the technical anonymisation measures.

The Commission's measures also aim to ensure that these controls remain in place over time, not only at the point of access. An OSE must obtain an independent reasonable assurance report (Level 1) before receiving the data and must then submit annual Level 2 reports, the first within six months of the Level 1 report, to show that the controls operate effectively in practice. The Decision also identifies a series of alerts that may signal a weakening of the controls: a late Level 2 report or one concluding only limited assurance, a change of control, insolvency proceedings, materially false information in the verification process, and a failure to comply with the erasure obligations. Some of these alerts trigger further assurance (a new Level 1 report after a change of control, an updated Level 2 report in case of insolvency, while others allow Alphabet to suspend or, in some cases, terminate access. The framework thus treats compliance as something to be verified continuously and not assumed once onboarding is complete.

These measures are crucial to the Commission's anonymisation claim: they allow it to argue that the residual risks of re-identification are reduced to an insignificant level. To be consistent with the EDPB's position, however, the Commission would also need to demonstrate that, with these contextual controls in place, residual risks are insignificant for all end users. However, it remains debatable, even under a contextual approach, whether these measures can genuinely be considered effective in addressing a given risk, particularly the risk of inference, where that risk has not first been mitigated through technical measures that can be assessed against a defined set of metrics.

## 3.4. Recall of Named Entity Recognition Tools

Alphabet is required to use personal data detectors to identify addresses, names, phone numbers, coordinates, IBANs, credit card numbers and other well-known identifiers as single entities to construct a allow-list of words that appear frequently. [3] Search queries that have entities that are not in the allow-list should be suppressed. Then under recital 609 the Commission notes that "Within the time limit set out in section 4.6.4, Alphabet must communicate for the Commission's review: (a) Which personal data detectors and which entity types Alphabet intends to use to recognise entities as part of query suppression, including the parameters such as the likelihood threshold (confidence) of each detector. Alphabet must explain why it has identified its proposal as the best performing option and why alternatives identified are less suitable."

Personal data detection is performed using Named Entity Recognition tools (NER). In practice such NER tools will not have a perfect recall and will miss some information. In the domain of detecting direct identifiers clinical notes, which is a high sensitivity type of document, a recall of approximately 0.95 is considered a practical performance benchmark for clinical-text de-identification. An earlier evaluation explicitly described 95% recall as acceptable performance for highly sensitive persona health information, including patient names and Social Security numbers [54]. More recent studies published between 2024 and 2026 provide additional benchmark context. Reported recall includes 95.7%–97.1% for an LLM-based system [54], 86% for a distilled language model [55], 95% for a transformer-based system [56], 95.55% for CliniDeID [57], 95% for an LLM-based system applied to large-scale Chinese EMR data [58], 98% for a deep-learning de-identification model [59], 96.25% micro-average recall for a full-span evaluation [60], 99.6% for a transformer-based ensemble [61], approximately 95% for pyDeID [62], and 98.05% for a locally deployed LLM-based anonymization approach [63].

Therefore, in practice there will always be a leakage rate. In the context of our current assessment, the risk model is based on the assumption that the controls and safeguards, as described above, will manage the residual risk and bring it down to an insignificant level. This is a very important component of a risk management framework for text that would be applied in the real world because otherwise no text identification would be acceptable as the recall values for these systems will not be zero. The question is what is the necessary recall rate for the personal data detectors here ?

We will consider as an example the situation of detecting a single direct identifier to compare to the allow-list. If a direct identifier in a query is not detected, then if that direct identifier is not in the allow-list, it means that the search query will not be suppressed when it should be. It would then be relatively easy to link that leaked direct identifier to a person (e.g., an end-user's name and identity).

[3] A caveat here is that this applies to signed-in end-users only. However direct identifiers can still be provided by end-users who are not signed in (e.g., self-searches) and can be about other individuals who are not the end-users.

We can compute the threshold recall for such a personal data detection tool based on our risk model. The re-identification basic risk threshold that we use in our analysis is 0.05,[4] which after accounting for controls and safeguards (see **Appendix A – Setting Single Team Vulnerability Threshold**, equation 5) gives us a personal data detection threshold of 0.1186. To meet that threshold, the tool would need to have a recall of **0.8814** if there is only one identifier in a query. That would ensure the direct identifier is detected and query suppressed if not in the allow-list such that the re-identification rate is below the threshold.[5]

Such a recall is achievable by contemporary tools. Even if we ignore the impact of the controls and safeguards in managing residual risk, the required recall of the tool would be 0.95, which is still consistent with what can be achieved today.

If we impose the additional requirement that matches with real individuals using the direct identifiers are affected by typical data quality issues and need to be verified (see **Appendix A – Setting Single Team Vulnerability Threshold**), then the required recall is **0.976**. This is a more challenging recall value to meet in practice and would imply that the standard required by the Commission is not very practical, especially when applied to heterogeneous data at scale.

The recall of the tool that was actually used during the anonymization is unknown to us. Although, the required recall value above would be possible to satisfy. In our analysis we will not consider attacks on such direct identifiers since they are almost certainly going to appear in the Search Dataset at a rate that is consistent with the tool's recall.

# 4. Motivated Intruder Test

## 4.1. Background on the Motivated Intruder Test

A motivated intruder test is an empirical test of data vulnerability. It typically does not consider the impact of additional safeguards that may be in place. However, in our study, the threshold that is used to interpret the results takes into account the impact of the safeguards.

A controller must consider all practical steps and means that are reasonably likely to be used by someone motivated to identify the people whose personal data was used to derive anonymous information, and that this motivated intruder test can be used to help assess the identifiability risk of apparently anonymous information [52]. A motivated intruder is defined as someone who wishes to identify a person from the anonymised information derived from their personal data, and the test assesses whether such a person is likely to succeed [52]. The intruder is assumed to be reasonably competent, to have access to appropriate resources such as the internet, libraries and public documents, and to use investigative techniques such as making enquiries of people who may hold additional knowledge about a person [52].

ISO/IEC 27559 adopts a comparable characterisation, describing a motivated adversary as relatively competent, with access to external resources and actively willing to make enquiries, while noting that a basic analysis does not assume specialist knowledge or recourse to criminality; it identifies adversarial testing using friendly adversaries as part of a practical assessment of a dataset's identifiability [9]. Such a test may be carried out by the controller, by a third party, or by both, and that an external organisation can contribute the perspective of an independent adversary together with awareness of data resources, techniques and vulnerabilities the controller may have overlooked.

[4] This eschews the assumption of five independent teams attacking the dataset as the data protection tools will be run only by Alphabet.

[5] This is only an approximation because some direct identifiers will be in the allow list and therefore would not result in the query being suppressed. Therefore, the rate of appearance of direct identifiers will be higher than the recall number suggests due to the high frequency direct identifiers being retained.

Published motivated intruder studies provide a model for how these tests can be conducted and re-identification outcomes and effort be reported. Branson et al. tested a clinical study report anonymised under European Medicines Agency Policy 0070, attempting to re-identify 500 patients in the United States. They reported six suspected matches, all of low confidence, at 170 hours in total and 24.2 hours per suspected match, with only social media and death records proving productive [64]. Their study also illustrates the importance of distinguishing suspected from verified matches and documenting effort spent on unsuccessful as well as successful approaches. For the Search Dataset, this reporting approach can inform the Commission's assessment of adversary effort.

In a penetration test commissioned by the Office for National Statistics (ONS), Elliot et al. [65] simulated an intruder with response knowledge attacking two social survey datasets, the Labour Force Survey (LFS) and the Living Costs and Food Survey (LCF), with 50 randomly selected target respondents each. Researchers spent on average about three hours of web searching per target in a first round and a further three to five hours on the 20 most visible targets, which we calculate to be 360 to 400 researcher-hours in total. They also bought commercial, people-search, and land-registry data, totalling about £13 spent per target. For the LFS, open information alone yielded 6 correct one-to-one matches among 50 targets (12%), rising to 12 (24%) when commercial data were added, or 14 (28%) including one-to-many matches. For the LCF, where the team held only addresses, the attack produced a single correct one-to-one match until a small-area geographic code was used; with it, 18 matches were correct, 14 of them one-to-one, from open information alone.

In the ONS intruder test of Census 2021 outputs [66], 30 of 51 recruited staff took part over three weeks, working in spare time around their regular duties, and made 81 identification claims, of which 40 (49%) were correct. Intruders reported typically taking 5 to 30 minutes to arrive at a claim, although total time could not be measured reliably, and most claims (59 of 81) concerned family and friends rather than strangers.

We conducted a motivated intruder test of the Search Dataset, within the ICO framework [52] and the adversarial testing approach of ISO/IEC 27559 [9]. An independent team attempted re-identification, more precisely identity disclosure, through three attacks, each an operational version of one of the Commission's criteria: a targeted attack beginning from a person, an untargeted attack beginning from records in the released data, and a scaled version of these attacks in which the online lookup was automated. The primary outcome was whether a mini-session could be successfully linked (and verified) to a named individual. For each attack, the number of matches and the effort required are reported.

## 4.2. Methods

### 4.2.1. Ethical Considerations

This project has been approved by the CHEO Research Institute Research Ethics Board protocol (REB Protocol No 26/127X). All methods were carried out in accordance with relevant institutional, national and international guidelines and regulations. The CHEO REB operates in compliance with, and is constituted in accordance with, the requirements of the Tri-Council Policy Statement: Ethical Conduct of Research Involving Humans (TCPS 2) [67]; the International Conference on Harmonization Good Clinical Practice Consolidated Guideline (ICH GCP) [68]; Part C, Division 5 of the Food and Drug Regulations [69]; Part 4 of the Natural Health Products Regulations [70]; and Part 3 of the Medical Devices Regulations [71] and the provisions of the Ontario Personal Health Information Protection Act (PHIPA 2004) and its applicable regulations [72].

Based on the conditions of the approval of the REB, we could not verify the identity of individual suspected matches between single individuals and end-users, and can only receive aggregate feedback on re-identification claims, without confirmation of any specific individual's identity. Acquaintance attacks were excluded to protect individual privacy and personal relationships (i.e., we could not attempt to re-identify friends and family).

### 4.2.2. Intruder Model

We make the assumption that the modelled intruder is an analyst at an OSE who holds the Search Dataset under Article 6(11) DMA and sets out to identify end-users in it. The Commission lists the recipient OSE among the adversaries it considers, alongside a malicious employee and an unintended third party [3] (recital 620), and describes OSEs as inherently capable actors with technical expertise, analytical tools and computational resources [3] (recital 830). In their motivated intruder test guidance, the ICO treats release to a defined group differently from public release: the assessment considers the knowledge and technical means available to that group, and an intruder may be someone permitted to access the data. The intruder's goal is to attach a name to a record (i.e., a mini-session) and, where possible, to learn something about that person not already known, which is the information gain the Commission weighs [3] (recital 624).

We used the reasonably competent intruder described in the ICO and ISO/IEC 27559 guidance as a starting point: an actor who uses public resources, has no specialist knowledge and does not resort to crime [9,52]. The ICO also recommends considering stronger capabilities, tools and resources when assessing high-value data. To reflect the technical capabilities of a recipient OSE, we extended this baseline in three respects.

First, the intruder knows the anonymisation measures and the data schema, and can enumerate bin sizes from metadata alone. Second, the intruder can use current LLMs and write SQL and Python. Third, the test draws on a multilingual team working over a three-week period with the efforts described in the section Time and effort. The tools available are described in the section Data Environment. These resources define the conditions under which we assess the effort required, using the Commission's dimensions of time, cost, manpower and technology [3] (recital 620).

The intruder's prior knowledge differs by attack direction. In Population-to-Sample (P2S) attacks, the intruder starts from a specific person not personally known to them, using a profile assembled from public sources: country and sometimes NUTS 3 region, language, a candidate date range from when the person was active online, and a distinctive interest or condition. The intruder knows neither the person's device type nor whether they appear in the Search Dataset. Individual matches cannot be checked against ground truth during testing; verification is described under the section Success Criterion and Adjudication of Suspected Matches. In Sample-to-Population (S2P) attacks, the intruder has no preselected target or prior knowledge of the individual concerned. They first select a search record or mini-session from the Search Dataset, then use its content and metadata to search public sources, seeking to link it to a named individual.

In the ICO's terms [52], the directions of attack differ in whether a specific or random person is targeted and in the background knowledge held. We did not assess re-identification attempts involving acquaintances, i.e., individuals personally known to the adversary, which the ICO recognises as a route to re-identification. Such attacks were excluded because an attack that depends on knowing the target personally cannot be repeated against a large number of end users, and so would not speak to criterion (c) and has additional ethical considerations (see section Ethical Considerations).

For both attack directions, external information was obtained through publicly accessible sources, manually or using LLM-assisted web search. Intruders did not join private groups, create accounts, use paid-membership services, make enquiries or contact individuals. This makes the model narrower than the ICO's, whose intruder makes such enquiries [52].

The model departs from a real recipient in ways that set the direction in which results should be read. The motivated intruder worked in an environment provided by Alphabet and was not bound by the contractual controls that apply to recipients: segregation, prohibitions on re-identification, linking and augmentation, and logging and audit [3]. The intruder faced no detection and no consequence for attempting what a recipient is forbidden to attempt, so results may overstate what a contractually bound recipient would attempt. Conversely, the environment's technical limits (no internet access, no coding assistant, copy-paste capped at a few hundred characters) are not assumed to bind a recipient, whose access is to be governed by licence and audit, and the arrangements for that access were still being settled when the study was designed. The intruder also held no auxiliary data beyond publicly available information, in particular not the recipient's own search logs and advertising data. The Commission says linking risk depends on the availability of such data [3] (recital 622(b)) and addresses it through contractually-imposed and audited logical segregation and prohibitions [3] (recitals 820–821). Specialist or criminal means, which maybe more appropriate for high-value data [52], were not modelled, and the team's motivation was demonstration rather than the commercial motivation attributed to the modelled intruder. The results therefore measure what the technical measures withstood in the hands of a capable but constrained intruder, not the risk presented by the sharing regime as a whole.

#### 4.2.3. Data Environment

The data was accessed within RIDGE, a locked-down virtual machine accessed by remote desktop and provided by Alphabet for the purposes of this study. While team members could import documents into the RIDGE environment freely, data export from the RIDGE environment was tightly restricted and no records or verbatim extracts from records were exported. There was no internet access inside the environment. API access to Gemini models was available within the environment and used as a tool in the motivated intruder testing.

We accessed the Search Dataset exclusively through SQL queries in Google BigQuery. These queries were subject to a daily scanning quota of 800 TB. Usage was measured by the volume of data scanned in the columns queried, rather than the number of records returned. For six months of German data, scanning metadata alone used approximately 4 TB; including query text increased this to 7 TB, click and interaction data to 14 TB, and result URLs to 105 TB. Scanning all 19 columns used approximately 350 TB, or 44% of the daily quota. Restricting searches to specific days reduced the volume scanned approximately proportionally. Consequently, broad searches could consume substantial resources even when they returned few matching records.

#### 4.2.4. Intruder team and study conduct

The motivated intruder team included 6 individuals with language skills spanning English, Estonian, French, Spanish, Turkish and German. Alphabet funded the study, provided the data environment and separately verified submitted matches, but had no role in designing the study, selecting attack strategies or conducting re-identification attempts. No intruder team member was affiliated with Alphabet.

The team members developed attack strategies, gathered publicly available background information and conducted P2S and S2P attacks using the tools available within the analysis environment, including the provided LLMs and installed Python libraries. The team worked collaboratively, sharing strategies and promising approaches, with tasks divided according to expertise, technical background and availability.

Testing ran from 19 October to 9 November 2026. In the week preceding the start of testing, all team members attended an onboarding session covering the project, the anonymisation measures and the rationale for the attacks and received instructions on environment access, time tracking, counting of targeted individuals and suspected-match reporting.

### 4.2.5. Attack procedures

The motivated intruder testing comprised three attack approaches: P2S, S2P and scaling through automation, addressing the Commission's residual risk criteria as summarized in **Table 2**. P2S starts from a person identified in public sources and searches for their records; S2P starts from selected search records and seeks to identify the person behind them. Automation extends these approaches to assess the potential for scaling. Intruders followed the procedures summarised in **Table 2**, while retaining discretion to adapt individual steps using the tools available within the study environment. Public-source searches generally preceded data retrieval in P2S and followed record selection and analysis in S2P.

**Table 2. Attacks and their correspondence to the Commission's residual-risk criteria**. S2P: sample-to-population; P2S: population-to-sample.

| Criterion | Commission's wording | Corresponding attack | What the adversary must achieve |
|---|---|---|---|
| (a) | An adversarial entity can choose a specific individual and, with significant probability, retrieve some of their search records with high certainty. | P2S | Start from a person identified in public sources and attempt to locate search records belonging to them. |
| (b) | An adversarial entity can choose a random search record and, with significant probability, re-identify the end user with high certainty. | S2P | Start from selected records within a small metadata bucket group and attempt to link them to a named individual, optionally linking multiple mini-sessions first. |
| (c) | An adversarial entity can perform attacks that re-identify a large number of end users with high certainty and with reasonably simple methods. | Scaling through automation | Perform either of the above at scale by automating the attack procedures and minimizing the effort required. |

**Table 3** summarizes the operationalization of the two attack directions. Intruders could adapt individual steps using the tools available within the study environment. In any attack, identifying named individuals relied on publicly accessible information and tools. Intruders did not access social media content requiring a logged-in account.

**Table 3. Operationalizing S2P and P2S Attacks.**

| Step | P2S | S2P |
|---|---|---|
| 1 | Outside RIDGE: A person or group identified from public sources | Within RIDGE: SQL retrieval from small metadata buckets defined by location, language, timeframe and device type, with preference for combinations that remain small across days |
| 2 | Within RIDGE: SQL retrieval from metadata buckets selected by location, language and | Within RIDGE: Typically use an LLM to filter for potentially identifying content and, where |
| | timeframe using publicly available information about the target; device type generally unrestricted | applicable, cluster mini-sessions believed to concern the same user (re-sessionisation), before manual review |
| 3 | Within RIDGE: Typically use an LLM to filter retrieved records for consistency with target profile before manual review | Outside RIDGE: Search public sources to link selected records or clusters to a named individual |
| 4 | Within RIDGE: Propose a link between one or more selected records and a named individual | |

### 4.2.5.1. Population-to-Sample

P2S targets were identified through public sources providing a name, location, relevant personal details and a plausible date range. Sources included event programmes, professional profiles and information about public-facing individuals. This online lookup (see details in **Appendix A – Supplemental Methods**) produced a target table containing each person's name and a confirmed location in one of the six countries covered by the study during the period for which search data were available. Profiles were then enriched with publicly available professional and personal details, such as country and NUTS 3 region, language of a public biography, a date window of up to one week taken from the event programme, professional position, employment history, education, languages, skills, hometown, hobbies and travel details such as conference attendance.

Target selection followed five general strategies: (i) event participants, particularly conference attendees expected to travel; (ii) public figures with extensive publicly available biographical information; (iii) individuals associated with sporting events, ranging from public figures to less publicly known participants; (iv) professionals, including lawyers and doctors; and (v) a refined strategy of (i) with travellers whose searches might concern their home location, institution or employer, including institutional and employer logins. These strategies sought to identify individuals whose activities, locations and backgrounds could provide sufficiently distinctive information to support linkage to search records.

Event-based targets were drawn from public programmes of conferences, congresses and festivals held between March and August 2026 in the six EU countries studied, and covered a wide range of fields. Roughly 40% of event-based targets were speakers or authors at computer science, artificial intelligence, software and cybersecurity events, including developer conferences and academic venues. About 20% were at medical and health congresses (for example oncology, paediatrics, palliative care and dentistry). The remainder were spread across the humanities, arts and literature (about 8%, such as book, film and music festivals), engineering and industry (5%), the natural sciences and mathematics (4%), business, finance and policy (4%), law (3%) and professional sport (3%, including marathons, cycling races and triathlons).

Target profiles guided retrieval from metadata buckets selected by location, language and timeframe, with device type generally unrestricted, as summarised in **Table 3**. Where useful, profile-derived terms were used to filter query text and clicked URLs. Retrieved records were assessed for consistency with the target profile, and proposed links between records and named individuals were submitted to Alphabet for verification. The tools used to support these steps and the progressive reduction in manual involvement are described in **Section 4.2.6.3**.

### 4.2.5.2. Sample-to-Population

S2P attacks began with records selected from the Search Dataset, without a preselected individual. Intruders used two selection strategies: (i) bucket-first, ranking metadata buckets by size using aggregate counts before examining query text, with preference for location-language-device combinations that remained small across days; and (ii) term-first, searching a country table for a potentially distinctive content term, such as a rare medical term, and selecting a metadata bucket from the results. Intruders chose the country, region, language, device type and date range.

Retrieved records were filtered for potentially identifying content. Where applicable, mini-sessions within the same metadata bucket were grouped into clusters believed to originate from the same person (re-sessionisation), based on query text and clicked URLs. Selected mini-sessions then informed public-source searches aimed at identifying the person behind them (see **Appendix A – Supplemental Methods**). Proposed links to named individuals were submitted to Alphabet for verification. The tools and human review used to support these steps are described in **Section 4.2.6.3**.

Records were deliberately selected rather than sampled uniformly at random, whereas the Commission's criterion (b) refers to a randomly selected search record [3] (recital 619(b)). The motivated intruder testing therefore evaluated re-identification of selected, potentially vulnerable records rather than estimating the probability of re-identifying an arbitrary record.

#### 4.2.5.3. Scaling through automation

All P2S and S2P attacks used Gemini to filter records and cluster mini-sessions believed to originate from the same person. Intruders initially reviewed the outputs and selected candidates for further investigation or submission to Alphabet. Automation was subsequently extended in P2S to reduce the human effort required for repeated re-identification attempts. The further automation focused primarily on event-based attacks because public event information facilitated automated target identification and profile construction. Travel to events helps distinguish individuals through searches reflecting their home location, employer or institution.

In P2S, automation extended to target identification and profile construction. LLM-assisted web searches using Claude or ChatGPT retrieved information from public sources, including conference programmes and professional directories, invoking Firecrawl where needed. The information was structured into target tables containing each person's name, available professional and personal details, inferred language, NUTS 3 region(s) and relevant date range. Target lists generally covered a single NUTS 3 region and a period of no more than one week. This process extended target identification beyond individuals selected through manual public-source searches.

The resulting target table was imported into RIDGE, where Gemini extracted potentially distinctive search terms from these profiles, including names, employers and topics. SQL queries then retrieved records matching those terms in the initial query, modified query or clicked-result URL, restricted to the relevant country, languages, regions and dates. Retrieved mini-sessions were passed to Gemini in batches of up to its input token limit, together with the target list. Gemini proposed links between mini-sessions and named individuals, returning supporting details, reasoning and confidence scores from 1 to 5.

This P2S workflow used two submission modes. With final manual review, intruders assessed Gemini's candidate links, assigned confidence scores from 1 to 5 and selected suspected matches for submission to Alphabet. Without final manual review, Gemini's candidate links were submitted directly to Alphabet for verification, using prompts refined through earlier manual reviews. Removing this review step further reduced human involvement between candidate generation and submission.

### 4.2.6. Success criterion and adjudication of suspected matches

The motivated intruder testing defined successful re-identification as correctly linking a search record or mini-session to a named individual. This criterion applied to both P2S and S2P attacks.

A suspected match linked one named individual to one or more mini-sessions. It became a suspected match when accepted after manual review or forwarded without review through the automated pipeline.

A submission was a suspected match sent to Alphabet for adjudication. Each submission named only one individual; where several individuals appeared plausible, the intruder selected the most likely candidate. Manually reviewed suspected matches received a confidence score from 1 to 5 in line with the scale adopted by Branson et al. [64].

Submissions were adjudicated in batches of at least ten. Alphabet returned aggregate results rather than decisions on individual matches. We therefore distinguish between suspected matches submitted by the team and aggregate adjudication outcomes; we do not report any individual suspected match as confirmed. Aggregate feedback was available during testing and could inform subsequent attempts.

Adjudication distinguished five categories: correct matches based on the name associated with the originating account; records attributed to a different individual; uncertain matches; records from signed-out users for whom no identity was available; and records that could not be adjudicated because of data-quality issues or user-requested deletion.

### 4.2.7. Analysis

For P2S, the targeted population comprised distinct named individuals for whom intruders retrieved and analysed records from the Search Dataset within RIDGE, whether or not a suspected match was produced. Individuals were deduplicated before processing. The number of targeted individuals formed the denominator for the targeted population re-identification rate. For S2P, we counted mini-sessions or clusters suspected to belong to one individual that intruders investigated through public-source searches to identify that person.

We recorded size of targeted population, suspected matches, time spent, including public-source research and unproductive retrievals, and Gemini input and output tokens consumed within RIDGE. We calculated re-identification metrics using three denominators: mini-sessions, submitted suspected matches, and the targeted population.

*Classification of outcomes.* We grouped the undetermined, signed-out and non-adjudicable categories as unclear. A submission was classified as correct (a hit) if at least one associated mini-session identifier was correct; unclear if none was correct and at least one was unclear; and incorrect otherwise.

*Unclear mini-session identifiers.* Alphabet could not adjudicate every mini-session identifier (see **Section 4.2.7**), so we report each mini-session-level and submission-level metric as a point estimate with a range. Our primary estimates exclude unclear mini-session identifiers, retaining submissions with at least one adjudicable identifier. We report a lower bound that counts unclear identifiers as incorrect; the upper bound excludes every submission with no correct identifier and at least one unclear one, mirroring the definition of a hit. Unclear mini-session identifiers counted as correct are reported in Appendix A as a further sensitivity analysis.

All relevant terminology and metrics for the motivated intruder attack are provided in **Table 4**. We assess re-identification success at both the mini-session and individual levels. Mini-session-level metrics count correctly attributed mini-sessions, while individual-level metrics count successfully identified people. Because several mini-sessions can belong to the same person, identifying multiple mini-sessions does not necessarily mean identifying multiple individuals. The targeted population re-identification rate counts each successfully identified person once and uses all distinct targeted individuals as its denominator, including those for whom no suspected match was submitted. Submission-level metrics describe the correctness of proposed matches, each linking one person to one or more mini-sessions.

*Population-normalised metric.* We also expressed the number of successfully identified individuals relative to the resident populations of the NUTS 3 regions examined, the countries examined and the EEA. Population figures were taken from Eurostat for 1 January 2025 using the NUTS 2024 classification [73].

*Stratification.* Analyses were stratified by country, NUTS 3 region, automation pipeline, attack strategy and, for automated attacks, manual-review status. Each factor was examined separately, without cross-tabulation. Results by NUTS 3 region and attack strategy were reported only for groups with at least ten submitted suspected matches. Adjudicated outcomes could not be compared between P2S and S2P because no S2P suspected matches were adjudicated by Alphabet.

*Time and effort.* We report testing hours for P2S, S2P, and strategy and planning, together with Gemini API tokens (input and output) consumed within RIDGE by attack direction. Hours per hit were calculated where adjudicated hits were available, both including and excluding strategy and planning time.

*Targets and suspected matches.* We report the number of targeted individuals, the number of mini-sessions associated with them, suspected-match counts and suspected matches per targeted individual.

*Confidence scores.* We report the distribution and mean of intruder-assigned confidence scores, overall and separately by country, NUTS 3 region, automation pipeline and attack direction. This analysis was restricted to manually reviewed candidates; Gemini-generated confidence scores for unreviewed candidates were not calibrated and were excluded.

*Potential information gain.* We descriptively summarised free-text notes accompanying manually reviewed suspected matches, supplemented by qualitative feedback from intruders, to characterise the additional information that proposed links might reveal if correct. This assessment was restricted to suspected matches and could not be linked to Alphabet's adjudication outcomes. Notes were not standardised, and their content and wording varied between reviewers.

### 4.2.8. Interpretation of adjudication results

In **Appendix A – Setting Single Team Vulnerability Threshold** we compute a threshold for data vulnerability (or residual risk). This threshold starts from existing precedents in standards and guidelines from regulators that have been explicit about acceptable risk. Our approach is conservative in that we take the lowest threshold in these standards and regulations given the volume of data affected, adjust it down further by accounting for common data quality issues and imperfect identity verification rates. We also make conservative assumptions about the impact of controls and safeguards in constraining the probability of attack.

The analysis in the appendix considers that there will be 5 independent teams that would be performing an attack on the dataset. These teams would be using similar strategies. This assumption again considers that an adversarial attack will be attractive to multiple actors and we would want to assess the overall risk to the data accordingly.

Based on that analysis, the residual risk threshold (data disclosure vulnerability in equation (1)) was determined to be 0.47% with a Monte Carlo 95% interval (0.42% - 0.54%). This threshold is based on the default impact of the controls and safeguards that have appeared in the literature. If we set a more conservative assumption about these controls (i.e., that half the time there will be an adversarial attack despite the controls) then the residual risk threshold is 0.32% (0.27% - 0.37%).

Consistent with the motivated intruder test that is conducted here, this threshold is defined for identity disclosure with verification of matches.

**Table 4.** Motivated intruder attack terminology and metrics.

| **Motivated Intruder Attack Terminology** | |
|---|---|
| Candidate match | A proposed association between one named individual and one or more mini-sessions, typically generated by GEMINI before thorough manual review. |
| Correct mini-session | A mini-session for which Alphabet's adjudication supports the proposed association with the named individual |
| Hit | A submission containing at least one correct mini-session. Other mini-sessions within the same submission may be incorrect or unclear. Also termed a verified submission. |
| P2S targeted population | Distinct named individuals (i.e., targets) for whom intruders retrieved and analysed records within RIDGE, whether or not a suspected match was produced. |
| S2P targeted population | Number of mini-session or clusters believed to originate from one individual and investigated through public-source searches to identify that person. |
| Submission | A suspected match sent to Alphabet for adjudication, associating one named individual with one or more mini-sessions. |
| Suspected match | A candidate match a team member accepted, or, with no review, one sent to Alphabet. |
| **Motivated Intruder Attack Metrics** | |
| Hours per hit | Testing hours divided by the number of hits. |
| Mini-session re-identification rate | Micro-averaged precision: Correct mini-sessions divided by submitted mini-sessions, with unclear outcomes handled as specified in the analysis. Also termed mini-session precision. |
| Population-normalised re-identification rate | Hits divided by the resident population of the specified geographical area. |
| Submission re-identification rate | Hits divided by submissions, with unclear outcomes handled as specified in the analysis. Also termed submission precision. |
| Suspected match rate | Number of suspected matches divided by the size of the targeted population. |
| Targeted population re-identification rate | The number of distinct targeted individuals with at least one mini-session adjudicated as correct, divided by the size of the targeted population. |

## 4.3. Results

### 4.3.1. Description of the Search Dataset

The present motivated intruder test accessed a subset of the Search Dataset covering 1 March to 31 August 2026 for six countries: France, Spain, Poland, Germany, Ireland, and Belgium. Data was organised into one table per country, partitioned by date. A separate table covering all 30 countries (the EU27, Norway, Iceland and Liechtenstein) was provided for August 2026 and supported summaries across the full geographical coverage. The August all-country table contained 58,163,659,270 records in 44,722,490,332 mini-sessions, averaging 1.30 records per mini-session, with a maximum of three by design. It totalled 353.73 TB and encompassed 65 query languages and 1,177 NUTS3 regions, with a median of 1.86 billion records per day.

Each record contained the initial and, where applicable, modified query text; displayed search results and their links, types and ranks; click and interaction data; and metadata describing the day, inferred language, location and device type. Interface fields included scroll count and the tabs shown and selected. Search results accounted for approximately 96% of the table's size.

The Commission's metadata buckets group records by day, inferred language, location and device type. Each bucket must represent at least 1,000 distinct signed-in users. In August, we observed 377,085 such buckets across all 30 countries, the smallest containing 1,245 mini-sessions. Overall, 95.2% of mini-sessions belonged to buckets containing at least 25,000 mini-sessions; these accounted for 35.7% of buckets. These observed sizes count mini-sessions rather than distinct users and therefore do not directly verify the user-count threshold.

Records were concentrated in a few languages: six accounted for 75% of records, while 37 out of 65 languages each accounted for less than 0.1%. Most searches occurred on mobile devices (about 81%), followed by desktop or laptop (about 17%) and tablet (about 2%).

### 4.3.2. Targets and suspected matches

During the motivated intruder testing, P2S attacks produced 1,972 suspected matches from 98,501 targeted individuals (2.0%). S2P attacks produced two suspected matches from 24 mini-sessions or clusters, each believed to originate from a single individual; neither was submitted to Alphabet because the submission threshold of 10 suspected matches had not been reached. No S2P outcome was therefore adjudicated.

Suspected match rates varied by country, strategy and level of automation. Targeted individuals were concentrated in Germany (30,848), Spain (25,094) and France (19,048). Country suspected match rates ranged from 1.06% (Belgium) to 2.65% (France). Across all strategies, the suspected match rate was 0.45% with a final manual review and 2.86% without manual review. For event-based attacks, the corresponding rates were 0.46% (31,101 targeted individuals) and 8.07% (7,029 targeted individuals). Although attempts were conducted for professionals (1,768 targeted individuals) and famous individuals (125 targeted individuals), these strategies produced six and zero suspected matches, respectively, and neither reached the submission threshold of 10.

Intruders assigned confidence scores to manually reviewed suspected matches. The mean score was 3.0 out of 5, with 58 (33%) rated 4 or 5. Tables containing full breakdown of attempts and suspected matches are presented in **Appendix A – Supplemental Results.**

Potential information gain was assessed from notes and qualitative feedback for manually reviewed suspected matches. Examples included use of employer or university services, such as email, internal portals and transcript requests; travel plans inferred from flight searches; and residential services, such as healthcare providers and local libraries. Query content and clicked URLs also indicated browsing interests, collection-related purchase interests and repeated engagement with particular listings or websites.

### 4.3.3. Adjudication by Alphabet

The adjudication results cover the 1,972 P2S submissions; no S2P suspected matches were submitted to Alphabet.

Adjudication identified 93 P2S submissions with at least one correctly attributed mini-session (see **Table 5**). The submission re-identification rate was 5.7% [4.7%–7.7%], with the range reflecting alternative treatments of unclear outcomes. The targeted population re-identification rate was 0.094% among 98,501 targeted individuals, treating unclear outcomes as unsuccessful identifications.

The 1,972 P2S submissions contained 6,878 mini-sessions, with a mean of 3.49 and a median of 1 per proposed individual match. Of these, 223 were correctly attributed (see **Table 6**), with a mean of 2.40 and a median of 1 correctly attributed mini-sessions per successful individual match across the 93 successful matches. Mini-session precision was 4.7% among determinate outcomes (223/4,778), or 3.2% when unclear outcomes were treated as incorrect (223/6,878). Detailed classifications are presented in **Appendix A – Supplemental Results**.

**Table 5.** Counts of targeted individuals, submitted suspected matches and adjudication outcomes.

| | |
|---|---|
| Target population | 98,501 |
| Targets with suspected match | 1,972 |
| Hit (at least one mini-session identifier verified correct) | 93 |
| Unclear (no hit; at least one mini-session identifier undetermined, not signed in or not adjudicable) | 761 |
| Incorrect | 1,118 |
| Targets with suspected match excluding those without any determinate outcome | 1,620 |
| Targets with suspected match excluding those with any unclear outcome | 1,211 |

**Table 6.** Mini-session counts and adjudication outcomes.

| | |
|---|---|
| Mini-sessions submitted as part of suspected matches | 6,878 |
| Correctly attributed | 223 |
| A different individual | 4,555 |
| Undetermined | 296 |
| Not signed in (no identity available) | 1,238 |
| Not adjudicable (data quality issues or user-requested deletion) | 566 |
| A different individual | 4,555 |
| Mini-sessions submitted as part of suspected matches with determinate outcomes | 4,778 |

The submission re-identification rate was lower with extended automation without final manual review than in workflows retaining final manual review (see **Table 7**): 5.2% versus 11.3%; the targeted population re-identification rate was higher with extended automation: 0.122% versus 0.045%.

**Table 7.** Submission and targeted population re-identification rates by suspected match review strategy.

| Workflow | Targets | Hits | Suspected matches | Submission re-identificaton rate | Targeted population re-identification rate |
|---|---|---|---|---|---|
| Workflow with manual review | 35,516 | 16 | 141 [100, 173] | 11.3% [9.2%, 16.0%] | 0.045% |
| Extended automation workflow without manual review | 62,985 | 77 | 1,488 [1,111, 1,799] | 5.2% [4.3%, 6.9%] | 0.122% |

By country (see **Table 8**), the submission re-identification rate ranged from 1.7% [1.5%, 2.1%] (Belgium) to 6.6% [5.5%, 8.6%] (France), and the targeted population re-identification rate ranged from 0.016% (Belgium) to 0.147% (France). Results by NUTS 3 region, for the 30 regions with at least 10 submissions, are in **Appendix A —Supplemental Results.**

**Table 8**. Submission and targeted population re-identification rates by country..

| Country | Targeted individuals | Hits | Submissions | Submission re-ID rate | Targeted population re-ID rate |
|---|---|---|---|---|---|
| Germany | 30,846 | 30 | 470 [302, 597] | 6.4% [5.0%, 9.9%] | 0.097% |
| Spain | 25,090 | 21 | 415 [348, 483] | 5.1% [4.3%, 6.0%] | 0.084% |
| France | 19,041 | 28 | 426 [327, 505] | 6.6% [5.5%, 8.6%] | 0.147% |
| Poland | 10,013 | 6 | 107 [78, 142] | 5.6% [4.2%, 7.7%] | 0.060% |
| Ireland | 7,087 | 7 | 150 [108, 176] | 4.7% [4.0%, 6.5%] | 0.099% |
| Belgium | 6,424 | 1 | 60 [47, 68] | 1.7% [1.5%, 2.1%] | 0.016% |

### 4.3.4. Time and effort

Testing required 283 hours: 153 for P2S, 32 for S2P and 98 for strategy and planning. P2S required 1.6 hours per confirmed match, excluding strategy and planning. Across the full study, effort was 3.0 hours per confirmed match including strategy and planning, or 2.0 hours excluding it. Applying the 2025 EU average hourly labour cost of €34.90 to the recorded study effort gives an illustrative human-resource cost of approximately €105 per successful match, including strategy and planning.

Hours per confirmed match could not be calculated for S2P because no suspected matches were adjudicated. Gemini consumed approximately 591 million tokens within RIDGE; detailed token usage is reported in **Appendix A – Supplemental Results**.

# 5. Discussion

## 5.1. Summary of findings

This study contributes to the discussion on the effectiveness of the Commission's anonymisation measures for search query data by combining a critical assessment with motivated intruder testing.

### 5.1.1. Summary of critical assessment of commission measures

The critical assessment found comprehensive contextual controls and safeguards, but identified unresolved questions about protection for all end users and inference risks. Specifically, the focus of anonymisation on end-users ignores the risks to other individuals that may be the subject of the queries. Hence such mixed data may still be personal information, as acknowledged by the Commission itself, even if all of the technical anonymisation measures are implemented. Furthermore, while inference risks may be inadvertently managed, the anonymisation measures and the tests specified by the Commission do not explicitly consider or protect against inferences. Again, re-enforcing the conclusion that even if all measures are implemented they would not necessarily meet all of the criteria stipulated by the EPDB in their Guidelines.

The performance of the Named Entity Recognition tools that can be used to detect personal information about end-users for checking against the allow-list must meet at least a recall of 0.8814, which is achievable. A more strict recall threshold is 0.95 if we ignore the role of controls and safeguards in managing residual risk, which is still achievable. However, we do not know the recall value of the tools that were used on the Search Dataset, and therefore do expect that direct identifiers about end-users were not flagged for checking against the allow-list. [6]

### 5.1.2. Summary of re-identification risk from motivated intruder test

The intruder test operationalized the technical concept of identity disclosure by attempting to link search records to named individuals, providing evidence relevant to the EDPB's assessment of singling out and linkability. The P2S and automation approaches extended earlier testing by the Commission and Alphabet by starting from named individuals identified through public sources and examining whether this approach could be scaled with reduced manual involvement. It demonstrated residual identity-disclosure vulnerability, with a targeted population re-identification rate of 0.094% among 98,501 targeted individuals, most of whom were assessed through a largely automated workflow.

Comparing the 0.094% targeted population re-identification rate that we obtained to our pre-defined threshold, **the 0.094% is below the residual risk 0.47% threshold**. It is below the 95% Monte Carlo lower whisker of 0.42%, and under even more conservative assumptions around the impact of controls and safeguards with a threshold of 0.32% (0.27% - 0.37%), our targeted population re-identification rate is still lower than the threshold.

When we examine these results by individual country, while there is variation, they are still below the threshold for every country considered. If we consider unclear verification results as correct, then the targeted population re-identification rate would be 0.867% and above our threshold (see **Appendix A – Supplemental Results**, Table 12). However, that would arguably be an optimistic result that exaggerates the effectiveness of the attack.

[6] Not all direct identifiers will be about the end-users. However, for this analysis we are only interested in the end-users.

When considering the NUTS 3 regions, only one exceeded the threshold, Arr. Mechelen (0.671%). If we use the more conservative threshold, then the region of Granada would also be above that threshold.

Note that the threshold is calculated assuming that 5 teams will be launching an attack, and makes assumptions about the impact of controls and safeguards to reduce the risk to an insignificant amount (i.e., the 0.47% threshold when combined with the impact of controls on the probability of attack, leads to a threshold that reflects common definitions of low overall or insignificant risk).

The simulation in the appendix also shows that the number of teams needs to go up to 25 independent teams attacking the dataset for the data disclosure vulnerability to exceed the threshold. The 25 independent teams is not an implausible number, but is not very likely. Therefore, only a large scale attack by 25+ teams would mean that the risk of re-identification, as defined here, would exceed the threshold.

Another way of looking at these results is that the success rate of the adversarial attack on the targeted population would have to improve 5x in order to reach the threshold. This is a nontrivial lift in performance that would be needed, and at least using the current strategies is unlikely to materialize.

These findings inform the assessment of the Commission's criterion (a), concerning retrieval of a targeted individual's search records, and criterion (c), concerning identification at scale. However, uncertainty remained at two levels from the perspective of the adversary: most submitted matches were incorrect or unresolved, and where at least one mini-session was correctly linked to an individual, other mini-sessions attributed to that same person could belong to someone else. These limitations constrain what the findings establish about an intruder's ability to identify users and reconstruct their search activity with high certainty. This means that even if a mini-session could be attributed to an individual, others in the cluster could belong to someone else and it would not be possible for an intruder to determine that without a reliable means of verification.

### 5.1.3. Summary on effort and resources

The average time for a single successful attack was 3 hours with the equivalent cost of €105. If we exclude planning and strategy development, the effort per successful re-identification is only 1.6 hours. This is a smaller amount of time compared to the Novartis study [64], but is consistent with the time it took in other studies conducted by the Office of National Statistics in the UK [65,66]. Note that our effort numbers do not include the effort that the Alphabet team spent adjudicating the suspected matches that were sent to them, therefore our total effort will be higher, and they do not include the token costs that were utilized with Gemini.

Arguably this amount of effort is not high, and a motivated intruder would not be deterred by this level of effort.

### 5.1.4. The importance of verification of suspected matches

The precision of our attack was 4.7%. This means that out of the suspected matches that were submitted, 4.7% of them were confirmed correct. Also, 38.6% of the suspected matches that were submitted for adjudication could not be verified either way, creating uncertainty on their status.

The important observation here pertains to the importance of the verification step. Suspected matches do not necessarily represent actual matches. In our case, if we considered the precision rate as the re-identification rate it would be much higher than the actual verified re-identification rate (4.7% vs 0.094%), and if we used the proportion that we believed were correct matches versus the total number of individuals actually targeted, the re-identification rate would also be much higher than the actual verified re-identification rate (2% vs 0.094%).

### 5.1.5. Key takeaways

Our analysis and results indicate that the Commission measures do not address the EDPB criteria fully, and that there are gaps. This means that even if the anonymisation measures met all of the stipulations made by the Commission, this will not ensure that the data should be considered anonymous according to the EPDB criteria. Part of this conclusion is derived from the lack of clarity in how to interpret some of the criteria in the EPDB Guidelines. But also part of it derives from the exclusion of non-end-users from the anonymisation measures, which means that there will likely be personal information for the non-end-users in the data.

Furthermore, the leakage of direct identifiers about end-users is expected to occur in practice because personal information detection tools do not, in general, perform at a perfect level. But the recall values needed to meet our thresholds are achievable using contemporary tools.[7]

On the other hand, our large scale motivated intruder test found that the re-identification rate relative to the targeted population is below a conservatively defined threshold. This means that for the purpose of ensuring low identity disclosure, the anonymisation measures have achieved that objective, within the constraints of our study. Even then, an adversary would not be easily able to verify matches against a ground truth and therefore would not be able to determine which suspected matches of individuals to search queries are true matches, creating nontrivial uncertainty on the side of the adversary in interpreting the outcomes of an attack.

Furthermore, the effort per successful match was low at 3 hours on average. This would likely not present a deterrent from an adversary to launch an attack on this type of dataset, especially if the motive was to discredit an organisation.

### 5.1.6. The counterfactual

If we suspend the facts and current risk model for a moment, and consider an alternative interpretation of our results whereby the contextual controls and safeguards stipulated by the Commission are given zero weight in the analysis. A conservative threshold that is used in practice is 0.05. The re-identification rate for the targeted population is still below the threshold. Therefore, the identity disclosure risk conclusion would not change. However, without the controls the rate of acceptable re-identification would be quite high and would impact a large number of individuals. Hence the real importance of the controls and safeguards as part of the risk analysis for such a dataset.

## 5.2. Automation and Re-identification Effort

P2S attacks produced all adjudicated suspected match submissions in this study. LLMs supported target identification and profile enrichment using public sources, proposed clustering of mini-sessions through re-sessionisation and matching against target profiles. Earlier workflows retained final manual review; later workflows further refined the workflow and omitted this step. The extended automation workflow generated 91.2% of submissions. Automated processing could run overnight without continuous human involvement, although some manual steps remained because of restrictions in the RIDGE environment (see **Section 4.2.6.3**). These remaining steps could potentially be automated under different access conditions.

The extended automation workflow achieved a higher targeted population re-identification rate (0.122% versus 0.045%), but lower submission precision (5.2% versus 11.3%). It therefore identified a higher proportion of targets while producing less reliable claims, increasing the verification burden per successful identification. A successful re-identification required only one correctly attributed mini-session; other mini-sessions attributed to that person could be incorrect or unresolved. Where clustering relied on content similarity, incorrectly attributed mini-sessions are expected to support similar information gain to correctly attributed records.

[7] We do not have data on the performance of the Alphabet tools, and this conclusion is based on results published in the literature.

As automation expanded, total study effort per successful match, including strategy and planning, fell from more than five hours in the earlier stages to approximately three hours overall (1.6 hours without strategy and planning time). This reduction coincided with growing experience, prompt refinement and changes in targets, so it cannot be attributed to automation alone. Automation reduced the need for human involvement, although growing experience, prompt refinement and changes in targets may also have contributed to the reduction in effort. Computational resources remained relevant: approximately 591 million Gemini tokens were consumed within RIDGE across all attacks.

The reported effort excludes independent verification of suspected matches, which Alphabet adjudicated in this study. Low submission precision would increase that additional burden, but does not by itself establish that re-identification requires disproportionate effort. Whether the effort is acceptable depends on the adversary's objectives, resources and expected information gain.

The feasibility of this workflow depends partly on the conditions of access. Effective segregation, restrictions on linking and augmentation, and monitoring of imports, processing and outputs could constrain the collection and matching of external information used here. Investment in obtaining and retaining access, together with consequences for non-compliance, may provide additional deterrence. However, recipient selection and investment alone do not establish low motivation among everyone with access to the data (see also [4]); protection depends on the practical effectiveness of the restrictions, monitoring and enforcement.

## 5.3. Re-Sessionisation, Record Isolation, Singling Out, Linkability and Identity Disclosure

Re-sessionisation focused on small groups of mini-sessions rather than extensive search histories. Among successful individual matches, the median number of correctly attributed mini-sessions was one (mean 2.40). Thus, successful identification did not typically establish a broader search history. This is compatible with the difficulties reported in the Commission's re-sessionisation testing [3] (recitals 754–758).

Difficulty reconstructing search histories does not preclude record isolation at the level of a single mini-session. Combinations of metadata, query text, clicked links and interaction data may distinguish observations within the released data, even where metadata alone place them within a large group. As discussed in the critical assessment, however, distinguishing a record does not by itself establish that its contributor can be identified.

Query content may provide clues that support identification. These clues may extend beyond the information retained in the generalised metadata. Previous research has shown that demographic characteristics, including age, gender and location, can be inferred with varying accuracy from query histories [30]. Where these characteristics are also known to an adversary, they may help narrow the set of plausible individuals. However, we could not validate such inferences from the released mini-sessions because we lacked independently verified demographic information about their contributors and could not contact them to obtain it. The decision also discusses Alphabet's argument that query text or result URLs can suggest a location more granular than the released location field. The Commission nevertheless emphasizes that such a clue does not establish which person within the metadata group issued the query [3] (recitals 681–684).

For example, an adversary may know that a person has travelled from a small hometown to a conference in another region and uses a minority language. Searches within the corresponding location, time and language group for residential services in that hometown, such as municipal waste collection or library opening hours, could support attribution to that person if they are the only plausible user with that combination of characteristics. Such attribution depends on evidence supporting that exclusivity and the assumption that the searches concern the searcher's own residence. Neither distinctiveness nor semantic consistency alone establishes a correct match.

These distinctions matter when interpreting singling out. As discussed in the critical assessment, its relationship with uniqueness and re-identification remains contested. We interpret singling out here as distinguishing a particular individual, taking account of the relevant population and information reasonably available to the adversary, rather than merely isolating a unique record. A further question is whether singling out also requires reliably associating the isolated records with that individual. If singling out under the EDPB guidance does not require this further association, the Commission's operational approach appears narrower: it requires that further association for successful re-identification [3] (recital 622(c)). If, however, the EDPB's requirement to go beyond record isolation implies such an association, the Commission's approach would appear consistent with that interpretation and with the technical concept of identity disclosure: correctly associating an individual with their records. Its three residual-risk criteria would then address this outcome and our motivated intruder test would operationalize singling out under this interpretation. The extent of alignment therefore depends on how the boundary between singling out and identification is understood.

Identity disclosure through singling out also depends on the performance of entity recognition and filtering. Missed or misclassified identifiers could remain in released query text, although we did not independently evaluate these systems. Research on text de-identification highlights the importance of assessing residual identification risk rather than relying solely on aggregate recall measures [74]. A retained name does not, however, establish the searcher's identity: the Commission found that self-searches were rare and names alone generally insufficient for attribution [3] (recitals 750–751). Additional query content, metadata or external information may help distinguish self-searches from searches concerning other people, although we did not separately assess this capability.

Linkability provides another route to identity disclosure. The Commission considers linking Search Dataset records with auxiliary datasets or contextual knowledge about an individual [3] (recital 622(b)). Its three residual-risk criteria assess the identification outcomes that such linkage may enable, rather than linkage alone. The motivated intruder testing in this study examined this pathway by operationalizing these residual-risk criteria.

### 5.4. Inferences and Information Gain

While identity disclosure, and the Commission's three residual-risk criteria, capture identification outcomes arising from singling out and linkability, inference requires separate consideration. We use the term "inferential disclosure" to focus on information learned about an individual beyond an appropriate prior-information baseline, rather than on any inference drawn from the data. This is consistent with the EDPB guidance's focus on specific and meaningful inferences about identified or identifiable individuals [4], as discussed in the critical assessment. Importantly, this does not necessarily require linking the individual to particular records.

Evaluating inferential disclosure requires assessing both the reliability of inferred information and what the released data adds beyond an appropriate baseline. Under the population-coverage assumption adopted here, membership is already expected, so learning membership alone would add little information; the remaining discussion therefore focuses on attribute disclosure.

Query content need not concern the searcher, and an inference may be possible even without that individual contributing to the dataset. The EDPB guidance expressly describes comparing inferences from two versions of the released data: one produced from the original dataset and another produced after excluding all records relating to the individual [4]. This approach also aligns with academic literature distinguishing information learned from an individual's contribution from information already inferable through background knowledge or population-level patterns [17,18,75]. Conducting such tests requires identifying the individual's contributions in the source data and rerunning the anonymisation procedure, capabilities unavailable under our access conditions.

The Commission's examples, however, primarily concern learning information after identifying an individual's records [3] (recital 622(a)). This addresses an important consequence of identity disclosure, but does not by itself assess attribute disclosure without record-level identification.

Information gain following identification is itself difficult to establish. In our motivated intruder testing, notes accompanying suspected matches suggested potential information gain concerning residential services, use of employer or university systems, travel and browsing interests. These observations could not be linked to adjudicated outcomes and therefore remain examples of potential disclosure. The significance of such information depends on the individual and context. Searches for residential services may reveal previously unknown aspects of a person's circumstances, while searches, clicks and interactions may disclose interests or activity. From the informational privacy perspective adopted here, any additional information correctly revealed about an individual is relevant, even if it appears mundane or does not concern special categories of data. Its significance depends on the individual and context. This relevance should be distinguished from whether a disclosure causes downstream harm, which a motivated intruder test is not designed to establish.

## 5.5. Ethical Constraints and Untested Attack Procedures

Motivated intruder testing raises ethical questions because it deliberately attempts to re-identify individuals and may reveal personal information. Consultation with an REB is therefore important, particularly where obtaining consent from targeted individuals is practically infeasible. In this study, further safeguards included aggregate adjudication, exclusion of acquaintance attacks, restrictions on exporting search content, no contact with individuals, and reliance on publicly accessible information, excluding sources requiring a social media login. These protections also constrained reporting, detailed analysis and refinement of attacks. The findings should therefore be understood as evidence obtained under an ethically bounded test, rather than an exhaustive assessment of all legal re-identification pathways with no constraints.

Acquaintance attacks warrant separate consideration because an adversary may already know contextual details that are unavailable publicly and could help recognise a person's searches. Restrictions on importing external information may offer less protection against knowledge already held by the adversary, although access controls, monitoring and consequences for misuse remain relevant.

## 5.6. Self-Reidentification and Social License

We did not test self-identification, where individuals recognise their own records using knowledge of their searches. This differs from re-identification by an adversary and may reveal little additional information to the individual. Nevertheless, self-reidentification demonstrations can attract publicity: in the UK Biobank example [76], a participant supplied personal information that journalists used to locate her record, according to UK Biobank's account.

Such self-re-identification attacks are also difficult to scale as they require each individual to provide specific details about themselves to facilitate an attack. Therefore, at best they can serve as demonstrations.

Such demonstrations can significantly affect public trust and the social licence for data sharing more broadly, even where they do not establish re-identification using information independently available to an adversary and ignore the protective effects of controls and safeguards. These potential consequences warrant consideration separately from what the demonstration establishes about anonymity.

## 5.7. Limitations

The findings apply to the selected target population, attack strategies and access conditions, rather than estimating re-identification probability for a uniformly random record or end user. Testing covered up to six months of data in six countries; longer observation periods, different auxiliary information or other attack strategies could produce different outcomes. As discussed above, the study environment and ethical safeguards constrained testing.

On the other hand, the targeted population were individuals with a good amount of publicly available information about them (e.g., lawyers will tend to have detailed biographies online). This makes them easier to match to search queries. Other members of the population may not have extensive publicly available information about them and therefore the re-identification risks for other targeted populations may be even lower.

Our effort calculation per successful re-identification did not account for the effort that the Alphabet team has spent on adjudicating matches. Therefore, effort is underestimated, although we do not believe that accounting for that effort would change the conclusion that the effort would not act as a deterrent.

Alphabet’s adjudication supported attribution to originating accounts, but could not establish who physically performed each search. Shared devices, shared accounts or inaccurate account identities could affect person-level attribution. Aggregate feedback also limited analysis of individual cases, as discussed above.

We did not independently evaluate the performance of the personal-information detection and entity-filtering systems or systematically target their potential errors. Their contribution to residual vulnerability therefore remains unquantified. We did also not assess whether malicious actors could manipulate entity frequencies in the source query data to cause otherwise excluded entities to meet the allow-list threshold. The feasibility of such manipulation, including the need to satisfy distinct-user requirements and bypass abuse-detection measures, remains plausible but untested.

# 6. Conclusions

Our analysis of re-identification risk considered the contextual approach to anonymisation described in the 2026 EDPB Guidelines and contemporary practices for assessing and managing re-identification risk that are consistent with the contextual approach. The analysis was both theoretical and empirical.

Our findings suggest that there are gaps in the Commission’s anonymisation measures that, even if implemented fully, would not necessarily ensure that the Search Dataset meets the criterion for being anonymous. The recall performance of the tools that would need to detect direct identifiers in the search queries to compare against the allow-list would have to meet certain thresholds, but will not be perfect and therefore there will be some leakage of direct identifiers. Given that the measures that were stipulated by the Commission addressed identity disclosure risks directly, by conducting a motivated intruder test we found that the identity disclosure risks are below a defensible threshold, and therefore would be considered insignificant.

It is important to distinguish between the simplified and the contextual approach, since the EDPB Guidelines 02/2026 treat them as two distinct approaches. The contextual approach is likely to do most of the work, even though a controller may choose the more restrictive simplified approach and decide to overestimate the likelihood that certain means will be used [4] (paragraph 48). Following the EDPB Guidelines 02/2026, the contextual approach makes it possible to meet the legal standard for anonymisation when the likelihood of re-identification is insignificant in reality.

In practice, this means that a series of security and privacy controls must be implemented at the data recipient, and these must be enforced through additional contractual controls. Such safeguards allow residual risks to be managed and treated as insignificant. However, the residual risks need to be quantified to be able to make that argument convincingly. Our analysis did so for a class of attacks and targeted populations.

Future work on this problem should investigate other targeted populations to quantify their re-identification rate (data vulnerability) and be able to generalize further the re-identification risk conclusions for the Search Dataset.

# 7. Role of Sponsor

This study was financially supported by Alphabet. The sponsor had no role in the study design; collection, analysis, or interpretation of the data; or preparation of the manuscript, with the exception of adjudication of suspected matches. The authors had full access to the search data as indicated in the manuscript and retained sole responsibility for the analysis, interpretation, and reporting of the findings.

# 8. Acknowledgements

In accordance with the Generative AI Delegation Taxonomy (2025), GenAI tools were used for proofreading and editing under full human supervision. The tools used were GPT-6.1 Sol and Opus 5.5. The authors take full responsibility for the content of the final manuscript. GenAI tools are not credited as authors and bear no responsibility for the final outcomes.

# Appendix

# A Critical Appraisal and Motivated Intruder Test of A Search Dataset Anonymised for the Purposes of the EU Digital Markets Act

# 1. Setting Single Team Vulnerability Threshold

The purpose of this analysis was to set a threshold for the targeted population re-identification rate for our motivated intruder team under realistic assumptions (the data vulnerability threshold). Here we formulate a model and run a simulation to compute that threshold. This model formulation was performed a priori before the motivated intruder testing exercise was conducted and therefore uses values and parameters that were determined from the literature.

While the main parameters can be derived analytically, the simulation allows us to consider additional parameters and assess variability under realistic conditions. Hence, it is additive to what we would obtain easily from an analytical exercise.

Our aim here is to compute a threshold for data disclosure vulnerability, which is the residual risk after the implementation of the transformation to the Search Dataset. We start from common quantitative definitions of insignificant risk based on precedents, and work backwards to define a threshold for residual or data disclosure vulnerability risk. That allows to evaluate us to interpret the motivated intruder test results of the targeted population re-identification rate.

## 1.1. Risk Model

The basic model we use is that an adversary will re-identify the data through two attacks: deliberate and a data breach. A third attack, an inadvertent re-identification is not considered.[1]

A deliberate attack is when an adversary deliberately attempts to re-identify an individual in the dataset. The risk probability of attack is determined by the controls. The default value for probability of attack using commonly defined controls and safeguards is 0.3 [1,2]. This value is intended to be conservative, and in our context the true probability of attack for organizations that receive the search data will almost certainly be below 0.3. Therefore, this value allows us to model a deliberate attack using a conservative assumption. The risk is therefore:

$$Risk_1 = V \times 0.3 \quad (1)$$

where $V$ is the vulnerability, which is the proportion of individuals that can be re-identified in the data.

For a data breach, we use the 0.126 value that has been computed for health data breaches [2]. This is informed by breach data reported to the US Department of Health and Human Services. We use this breach rate because it is, relatively speaking, a reliable number to use. The health number does not apply to search data, and covers many small-sized organizations that may not have had robust data protection practices. Following the same argument as above, we chose a value that is conservative to allow us to model the breach rate. Hence we have:

$$Risk_2 = V \times 0.126 \quad (2)$$

The probability of at least one of these two risks occurring is:

[1] Given the large volume of data, an inadvertent scenario whereby an analyst would spontaneously recognize someone that they know is unlikely. Manually browsing such a volume of data is not practical.

$$1-\left(1-0.3V\right)\left(1-0.126V\right) \quad (3)$$

Therefore, this overall risk value represents a conservative (i.e., inflated) value of risk. If the risk using the conservative value is low then it will also be low for the true value. If the risk using the conservative value is high, then one can argue that it is erring on the side of protecting the privacy of individuals.

We want to make sure that this probability is below a predefined threshold. Under existing standards and regulations for non-public data release, the most conservative threshold is 0.05 for sensitive information [1–3]. We chose the lowest recommended threshold because of the large volume of data, and hence the large number of individuals that could potentially be affected by a re-identification, as well as the potential harm due to the intimate nature of some of the search queries [1].

However, this value assumes that data and the adversary knowledge do not have quality problems and have no errors, hence matching a dataset with background adversary knowledge is error-free. This is unrealistic in the context of a motivated intruder test since we are working with real data. An end-user may search using incorrect information, information about someone else, for example.

Real data has errors in it and therefore the accuracy of the matching based on adversary knowledge will be reduced [4,5]. Known data error rates (e.g., in voter registration databases, surveys, and data from data brokers) can be relatively large [6–9]. For health data, the error rates have tended to be lower [10–14]. If we take a weighted mean of 5% error in each piece of information, then the probability of at least one piece of information having an error in it is given by $1-(1-0.05)^2$ is 10%. This assumes that the search data will have two pieces of information that can be used for re-identification, which is consistent with assumptions made in re-identification risk assessments of unstructured data [15]. If at least one piece of information has an error in it then the matching would fail.

In addition, not all candidate matches will be verified. A previous review of re-identification attempts found that when there is a suspected match between a record and a real individual, that suspected match could only be verified 23% of the time [16]. This means that only 23% of suspected match can plausibly be verified as correct. In our context verification can fail, for example, because a user was not logged in and therefore their identity cannot be determined.

Therefore, the probability of correct data that can be verified is 0.207. Therefore, we adjust the threshold by this number to get a threshold that is more consistent with a motivated intruder test where both data quality and unverified matching will have an impact. This gives us a threshold of 0.01 that we will use in this analysis.

Now, going back to equation (4), we can formulate this as:

$$1-\left(1-0.3V\right)\left(1-0.126V\right) \leq 0.01 \quad (5)$$

And therefore the threshold data vulnerability $V$ that we are aiming for is 0.0235.

## 1.2. Multiple Teams

In our motivated intruder test only one team performed an attack on the dataset. However, in practice there may be multiple teams that perform an attack on the dataset. If we make the assumption that multiple teams will perform an attack on the dataset, then we need to adjust the data vulnerability threshold. We do so by conducting a Monte Carlo simulation.

For the sake of our analysis we assume that there is a limit of five teams performing a re-identification attack. These can be five teams within the same data recipient or five teams across data recipients. We then set the 0.0235 threshold for an individual in the dataset to be detected by at least one of the five teams.

We further make the assumption that the different teams will follow somewhat similar strategies, and therefore the probability of re-identifying an individual by one team is correlated to the probability that the same individual will be re-identified by another team. This is not the same as the same individual (end-user) being re-identified by other teams, but only that the probability of detection is correlated.

If we have $N$ individuals in a dataset, we can model the probability that an individual is re-identified as a Beta distribution. We assume that all individuals in the dataset have the same re-identification Beta distribution. And to capture that these probabilities are correlated across motivated intruder testing teams, we set $r$ as the pairwise correlation of the Beta distributions for any individual between two teams. This value is the same across all individuals and team pairs. This is a Pearson correlation. Hence the data is represented as a Gaussian copula and then converted back to a Beta distribution. This means that the correlation on the Beta distribution will not be the same as the $r$ value in the Gaussian representation.

For the baseline simulation we set $N = 10000$. The sensitivity analysis below shows that the results are not sensitive to this parameter. By default we set the $r = 0.15$ to indicate weak correlations among the teams. However, as we will see below, the main results are not sensitive to this value.

We set the value of the Beta distribution such that the probability of an individual being re-identified by *at least one* team to 0.0235.

## 1.3. Simulation

To perform a simulation, we sample the $p_{ij}$ value 1000 times, which represents the probability of re-identification for individual $i$ by team $j$ from the $Beta(\alpha, \beta)$ distribution. Next, we draw $Y_{ij} \mid p_{ij} \sim \text{Bernoulli}(p_{ij})$. For each iteration we compute:

$$D_i = \frac{1}{N}\sum_{j=1}^{N} Y_{ij}, P_1 = \frac{1}{N}\sum_{j=1}^{N} \mathbf{1}\left\{\sum_{i=1}^{k} Y_{ij} > 0\right\}, P_2 = \frac{1}{k}\sum_{i=1}^{k} D_i.$$

where $D_i$ is the proportion of individuals that are re-identified by team $i$. Thus $P_1$ counts each detected individual once across teams. $P_2$ is the average team-specific proportion.

## 1.4. Results

The main results for the two parameters of interest at 5 teams are shown in **Figure 1**.

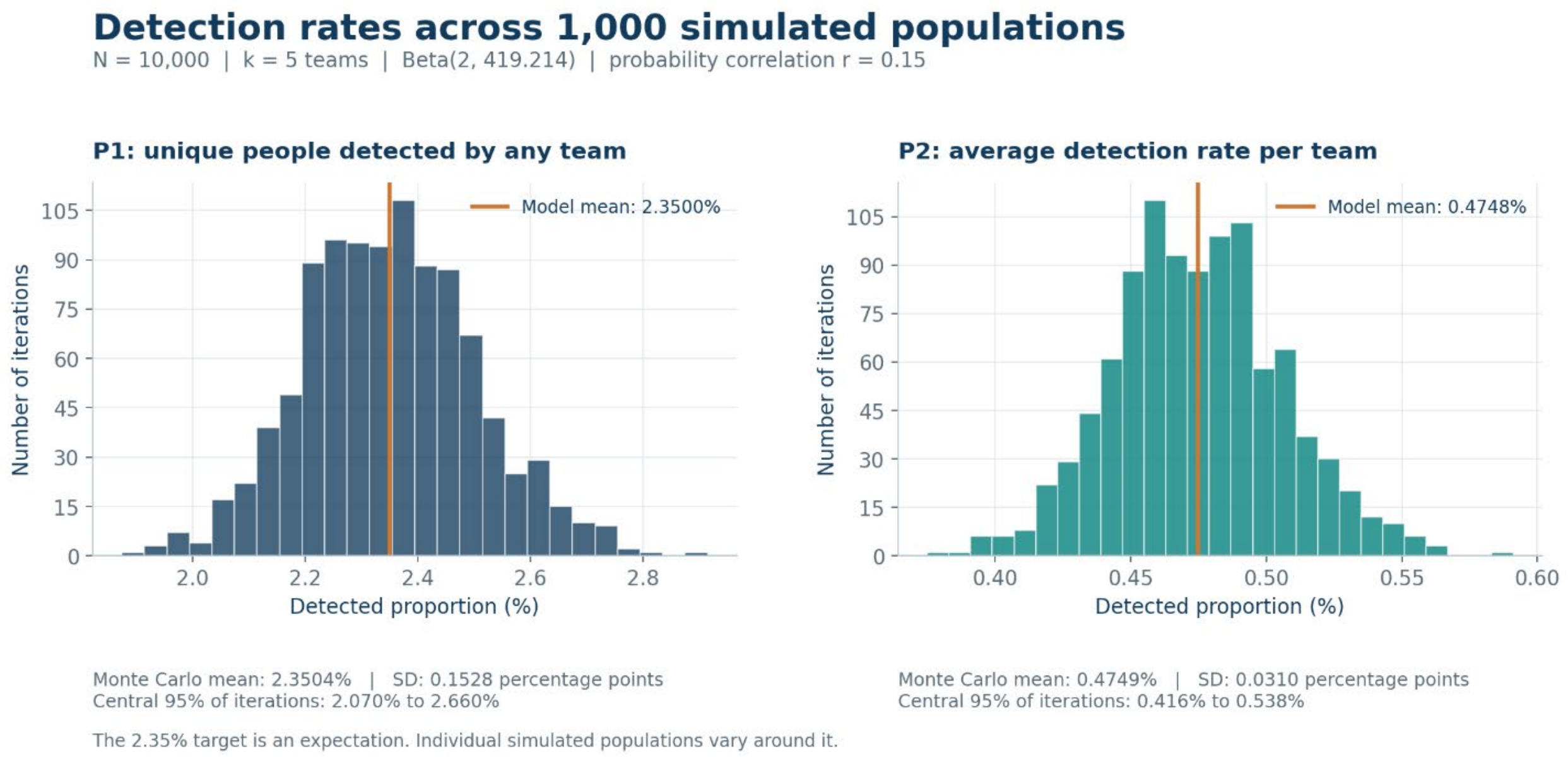

**Figure 1.** Distribution of the two parameters of interest across the simulation iterations.

We can see that the vulnerability of an individual across all teams is at our baseline value, and the mean per team re-identification rate on average (Monte Carlo 95% interval) is **0.0047 (0.00416, 0.00538)**, which sets our threshold for the vulnerability that would indicate that the risk is too high. Therefore, when the data vulnerability reported by a single team exceeds this threshold value, then the re-identification risk is deemed to be too high, and if it is at or below this value then the risk is deemed to be acceptably low. This threshold value is computed based on the assumptions that there are suitable controls and safeguards in place, that there are data errors, and that not all suspected matches can be verified.

## 1.5. Sensitivity Analysis

We evaluated the sensitivity of our results to the population size, the $N$ value, and the correlation, the $r$ value, as well as the number of teams. Each of these parameters was varied independently keeping each of the others at the default values. We can see from the results in **Figure 2** and **Figure 3** that the per team vulnerability is not sensitive to either of these values. Although larger population values result in less variability in the proportion re-identified values.

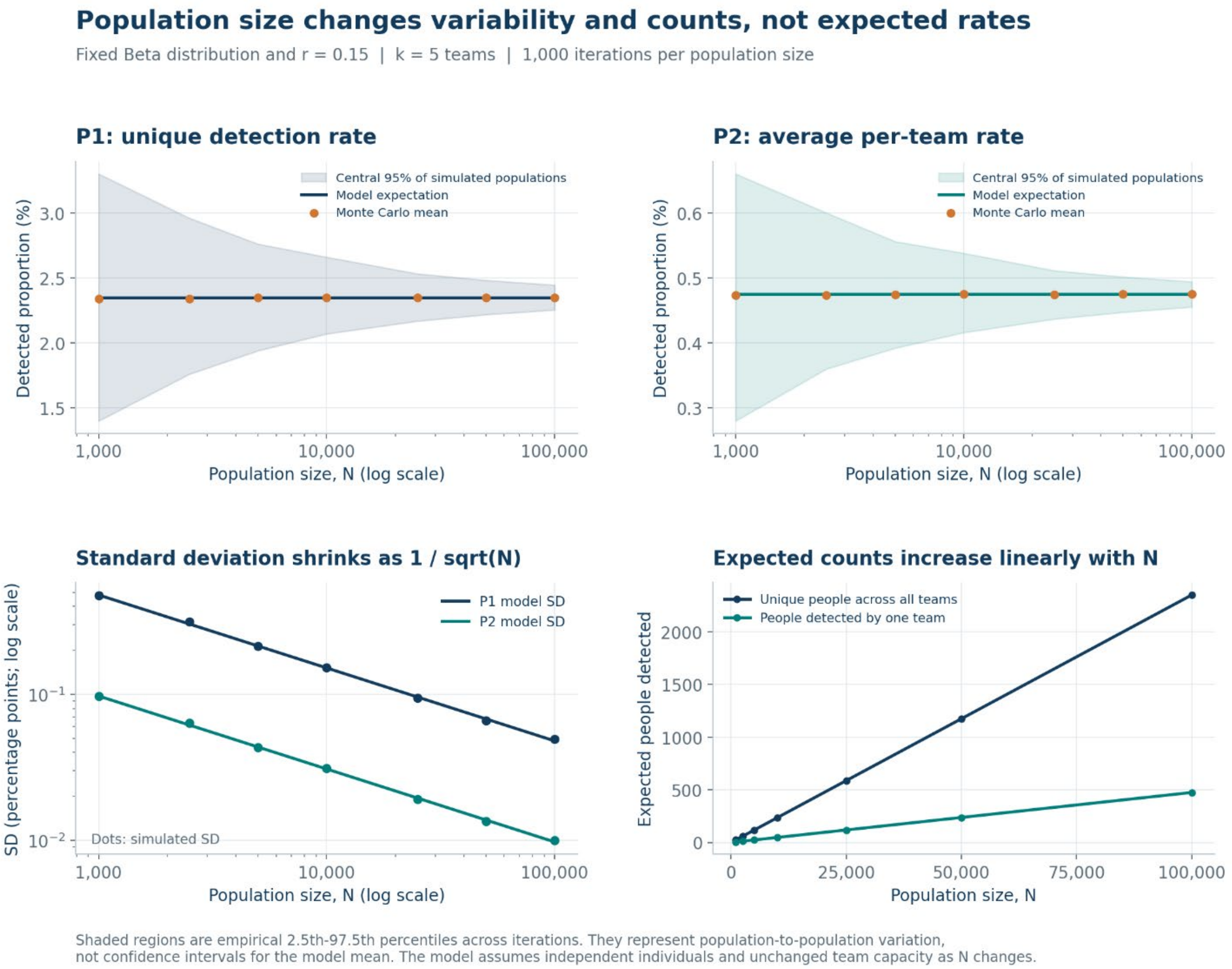

**Figure 2,** Sensitivity analysis of how changes to N have an impact on the P1 and P2 values.

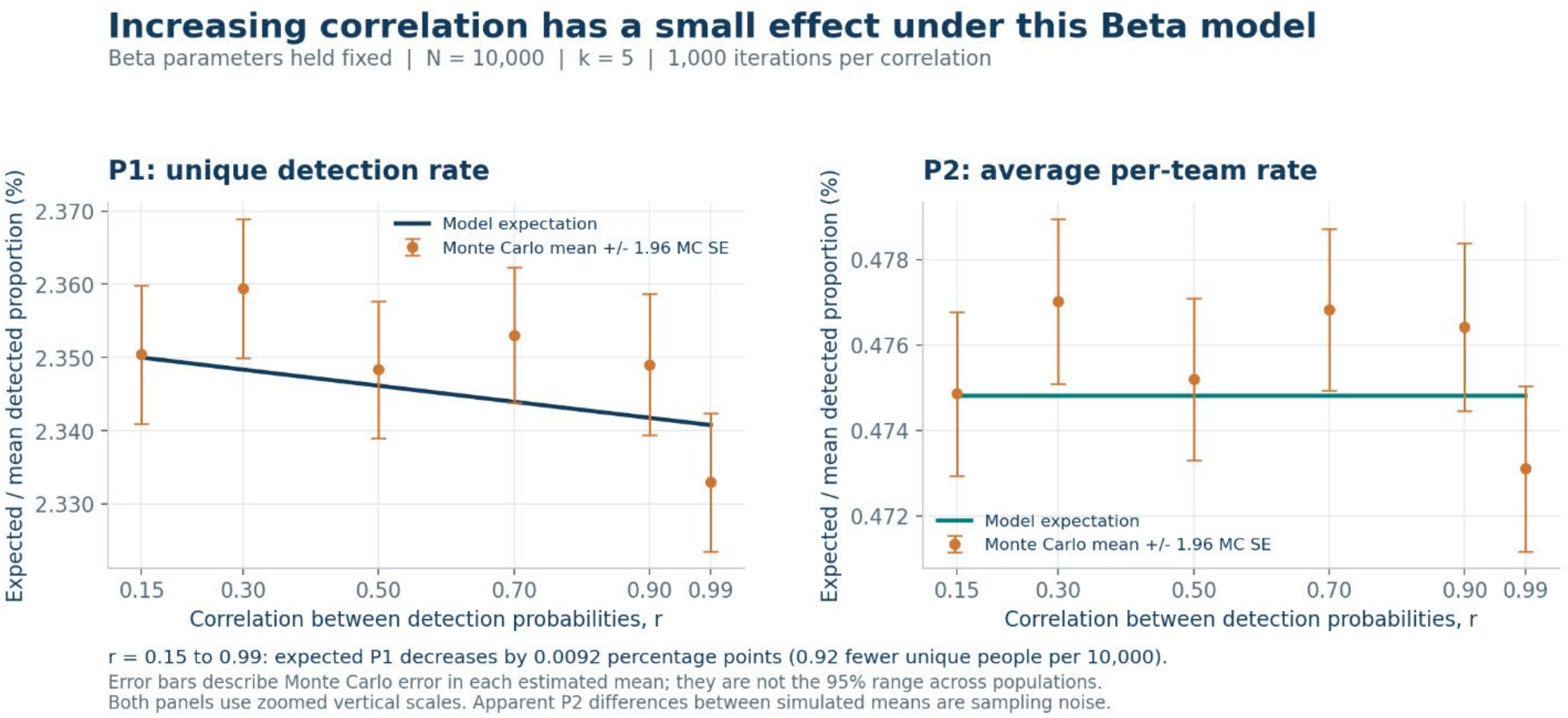

**Figure 3.** Sensitivity analysis of how changes to r have an impact on the P1 and P2 values.

We also examined the sensitivity of the results to the number of teams, as this was an important assumption, using the same default parameters. The results are shown in **Figure 4**.

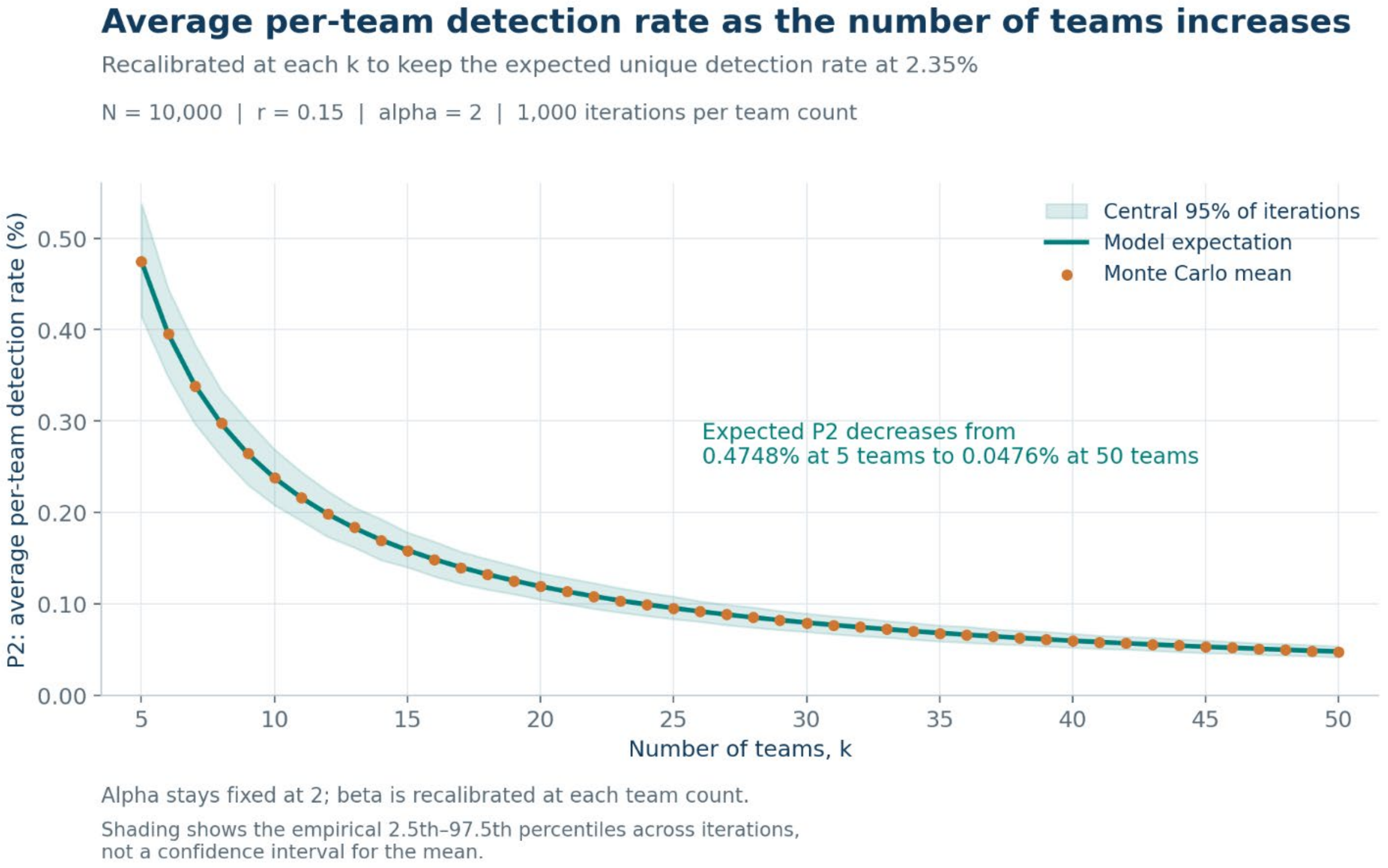

**Figure 4.** The impact of the number of teams on the threshold value.

We see that there is considerable sensitivity, as would be expected. The question then rests on what the plausible number of independent teams that can launch an attack on the dataset under the same strategies that were used in the current analysis.

Another item that we examined was the probability of attack value under the deliberate attempt. The default value used is 0.3, which is already quite conservative, has been used extensively in practice, and has been recommended in regulatory guidance. However, to evaluate the impact of changing the value, we changed that to 0.5, which means that 50% of the data recipients will attempt a deliberate attack with five independent teams. This is somewhat unlikely but allows us to examine the extremes of the distribution. These results are shown in **Figure 5**.

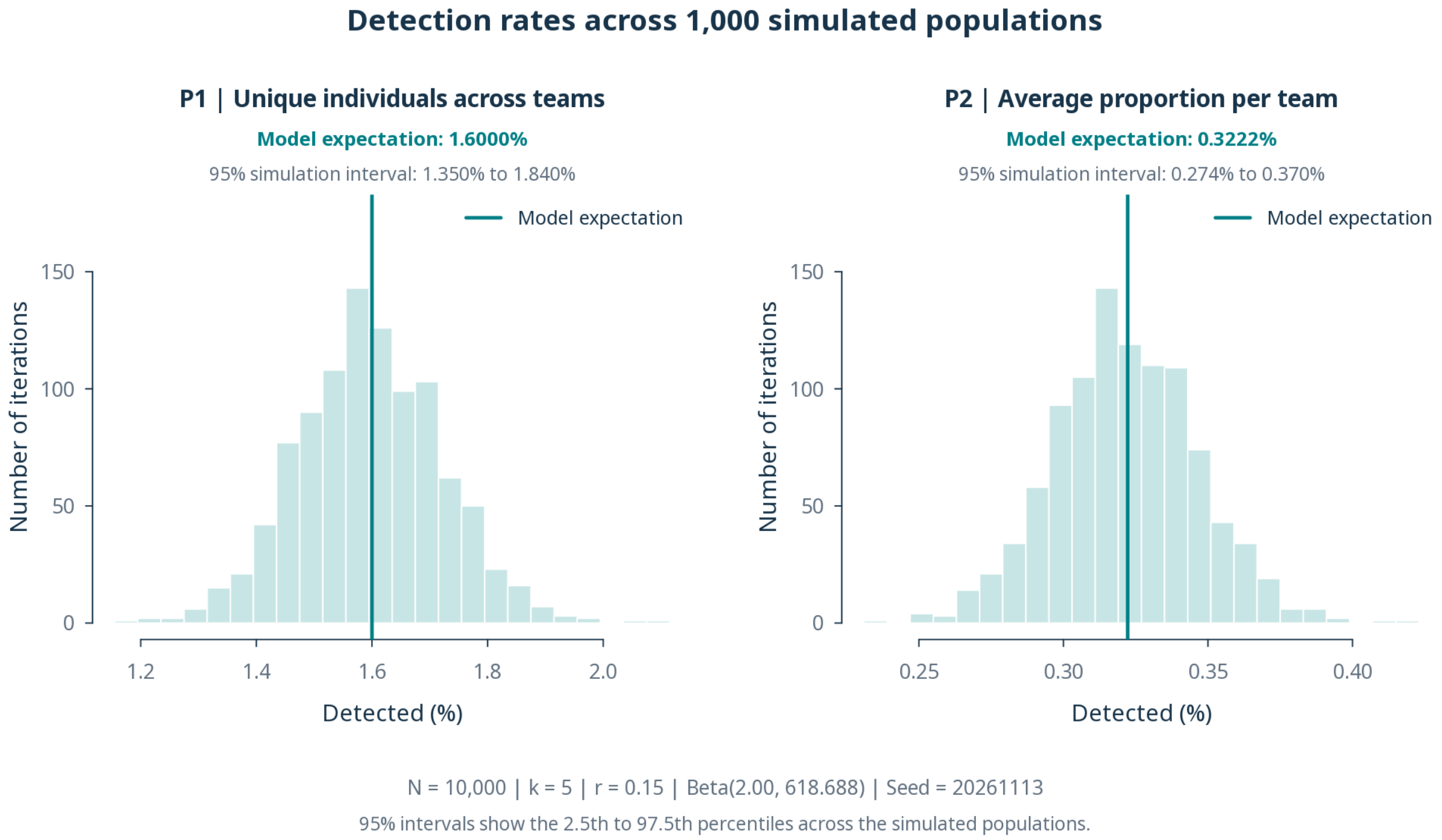

**Figure 5.** The proportion of individuals that would be re-identified with a probability of attack value changed to 0.5.

Under this set of assumptions, the threshold shifts to 0.0032 (0.00274,0.0037). This is a lower threshold than the one we considered above and can be used during the interpretation of the results to examine their robustness.

## 1.6. Limitations of Analysis

The main limitation in these results is that we assumed that five teams represent the most plausible scenario for an attack across the users of the data by default. Our reasoning is that if there were five teams that were attempting to re-identify the search data, then this would be detectable and would result in remedial action. Although the sensitivity analysis shows the impact with different team numbers.

Another limitation is that we assumed that the Beta distribution applies consistently across all regions. It has a larger mass at the low end, which is reasonable. Nevertheless, it may be difficult to re-identify individuals in more remote regions, for example, because there would be less known about them. Furthermore, we used specific strategies, and these may favour more urban regions.

### 1.7. Main Takeaways

This simulation established the **0.0047 (0.00416, 0.00538)** as the data vulnerability threshold that will be used for evaluating the results of the motivated intruder test. This value assumes that the contextual approach in the EDPB Guidelines is being followed, and hence that adequate controls and safeguards are in place. This number also assumes that five teams will be performing independent attacks using similar strategies to ours.

# 2. Supplemental Methods

### 2.1. Online Lookup

The online lookup was performed outside the analysis environment, since the environment had no internet access. Its role differs by attack direction. In P2S, online search came first and was exploratory, finding and profiling a target before any data were pulled. In S2P it was the final stage: where an intruder judged a reviewed cluster worth pursuing, they used summaries and extracted entities from the search query clusters—never raw data—to re-identify individuals via online searches.

Permitted sources were those the ICO identifies as available to a motivated intruder and reachable through an ordinary browser search: public records, genealogy sites, online services including social media, internet searches, local and national press archives, and generative AI tools [17]. Public social media was viewed without an account, sometimes using third-party viewer tools. Web-search tools, such as Firecrawl used through Claude connections, were permitted by the data controller when applied to abstracted information and target profiles, but not to verbatim anonymised data. Intruders did not join private groups, create accounts to gain access, use paid-membership services, or make enquiries of people who might hold additional knowledge about a person, and no individual was contacted at any stage. Targets were non-acquaintances identified from public sources. In these respects the test was more conservative than the ICO's model intruder [17].

The data controller set the terms on what from the released data could be entered into an external search tool, to contain leakage of the Search Dataset. Verbatim anonymised data, including query text and result URLs taken from a record, could be cross-referenced only through Alphabet tools; this was the threshold the controller set. Other tools, including the interactive search tools above, could be used only on abstracted information, such as a specific extracted entity (a name, place or handle), a public fact, a keyword, or a profile compiled from public sources, and never on full query strings or raw record content. The level of abstraction that enabled successful attacks while minimizing risk of raw Search Dataset being leaked was negotiated with the data controller.

# 3. Context Controls Comparison Analysis

The following is the analysis of the mapping of the controls and safeguards described in the Commission report and the Ontario de-identification guidance.

**Table 1:** Controls Coded from the Commission Implementing Decision.

| Number | Controls | Source |
|---|---|---|
| **Controls for ensuring anonymization from OSEs' perspective** | | |
| C1 | Effective control over the processing environment: The processing environment of the Search Dataset must be under the effective control of the OSE. Any processing of the Search Data, including processing performed by a subcontractor, must happen in the processing environment under the OSE's control. The processing environment can be hosted by a subcontractor acting as (sub)processor in the cloud but subject to the same restrictions. | Recital (791) (795) (810) (811) (819)<br>Annex (46)(c)(d)(e), (48) |
| C2 | Segregated processing environment: The processing environment must be logically and technically segregated. No physical separation of infrastructure is required. | Recitals (820), (837)<br>Annex (49) |
| C3 | Prohibition against linking, re-identification and augmentation | Recital (821)<br>Annex (50) |
| C4 | Purpose limitation: the Search Dataset shall solely be processed for the permitted purpose of optimizing and improving OSE services | Recital (822), (823), (824)<br>Annex (51), (52), (53) |
| C5 | Testing safeguard for ML models (pre-deployment evaluation of memorization of the Search Dataset) | Recital (825)<br>Annex (54) |
| C6 | Maximum retention duration: 13 months | Recital (826)<br>Annex (55) |
| C7 | Prohibition of onward sharing and disclosure to the exclusion of (sub)processors subject to same restrictions | Recital (827<br>Annex (56) |
| C8 | Governance and traceability measures to ensure and demonstrate compliance with measures imposed through contractual obligations. Obligation to document automated systems that access the data, material processing operations, and models trained or fine-tuned on it, with the results of the pre-deployment testing | Recital (828)<br>Annex (57) |
| **Controls for ensuring integrity and confidentiality and therefore anonymisation against any other person** | | |
| C9 | Technical protection of the Search Dataset via encryption at rest and in transit and appropriate cryptographic key management measures | Recital (912)<br>Annex (59) |
| C10 | Access control following need to know and least privilege principles and access governance including enforceable confidentiality obligations and yearly training. Phishing resistant multi-factor authentication is required | 913<br>Annex (60) |
| C11 | Data handling and other protection measures (prohibition of copying, export outside the segregated processing environment and access through other devices such as employee-managed devices or workstations) | Recital (914)<br>Annex (61) |
| C12 | Physical security measures to prevent unauthorized physical access to systems and storage media used to store or process the Search Dataset | Recital (915)<br>Annex (62) |
| C13 | Logging, monitoring and accountability measures to ensure traceability and accountability for all access to the Search Dataset. Access logs must be retained for 1 year and must be protected. | Recital (916)<br>Annex (63) |
| C14 | Data lifecycle management measures including erasure or irreversible deletion after expiration of data retention period. | Recital (917)<br>Annex (64) |
| C15 | System security maintenance and vulnerability management measures including proactive measures to detect and remediate vulnerabilities before they are exploited | Recital (918)<br>Annex (65) |
| C16 | Data breach reporting and liability (obligation to implement processes to detect, report and respond to data breaches) | Recitals (919-926)<br>Annex (46)(e), (66) |
| **Independent verification mechanism** | | |
| C17 | Obligation to produce Level 1 assurance report in accordance with ISAE 3000 providing reasonable assurance on the first Commission-defined assurance objective as well as on the design and suitability of the controls put in place to meet the second and third Commission-defined assurance objectives. Policies and actual implementation of controls should be checked. Submission of non-confidential level 1 assurance report necessary for access. The provision of materially false information is subject to financial penalties. | Recital (934-(939), 1150(b), (1195), (1202), (1326)<br>Annex (69), (72), (127), (129), (169) |
| C18 | Obligation to produce yearly Level 2 assurance reports in accordance with ISAE 3000 providing reasonable assurance on the first Commission-defined assurance objective as well as on the operating effectiveness of those controls and covering the Commission-defined assurance objectives. Policies and actual implementation of controls should be checked. The first Level 2 report is due within six months, then every 12 months. The Commission alone can require an ad hoc Level 2 report. Submission of non-confidential level 2 assurance report subject to access suspension and termination. The provision of materially false information is subject to financial penalties. | Recital (934-938), (940), (1222-1224), (1233), (1273-1274), (1279), (1285-1289), (1326)<br>Annex (73), (139-140), (142), (154-158), (169) |
| **Other controls** | | |
| C19 | Change-of-control notification obligation, with a new Level 1 report, subject to expedited access termination and financial penalties. | Recitals (1247-1248), (1311), (1318-1321), (1326), (1330), (1371)<br>Annex (144–147), (169) |
| C20 | Insolvency proceedings notification obligation subject to access suspension and termination and financial penalties. | Recitals (1248-1263), (1273), (1276), (1280-1281), (1326)<br>Annex (148-151),(154) (157), (169) |
| C21 | Transfer impact assessment if relevant to be performed by Alphabet prior to sharing | Recitals (134), (174), (183), (878) (1198), (1237)<br>Annex (9), (130) (146) |
| C22 | Information obligation imposed upon Alphabet to inform through its website data subjects about eligible data beneficiaries and the existence of level 1 and level 2 assurance reports and where relevant eventual termination of access to the Search Dataset. | Recitals (877), (1345)<br>Annex (176) |
| C23 | OSEs are independent data controllers for personal data within the Search Dataset and therefore subject to the whole full range of data protection principles. | See e.g., Recitals (551), (790), (791), (797), (800), (877), (878), (919)<br>Annex (45), (46) |
| C24 | Subcontractors must be bound by all the controls and must give auditors verification rights. | Recital (795)<br>Annex 46(d) |

**Table 2:** Mapping of Controls between EC Case DMA.100209 and Appendix A of Ontario Guidance.

| EC Case DMA.100209 | Appendix A (Ontario De-Identification Guidelines for Structured Data) |
|---|---|
| C1 | Minimising and specifying storage locations in advance, and organisational storage and access policies. See also Appendix C. |
| C2 | Special Protection for remote electronic access to data |
| C3 | Data user staff under a nondisclosure or confidentiality, or pledge of confidentiality, See also Appendix B and C. |
| C4 | Independent advisory or data stewardship committee oversees scope of data processing. See Appendix C. |
| C5 | |
| C6 | Data is securely disposed of after a specified retention period. Periodic audits of long-term retention, retention schedules, and see Appendix C |
| C7 | Appropriate controls and prior approval for disclosure or sharing of the data by the user with third parties |
| C8 | Organizational policies for storage, management and access<br>Organizational governance framework for privacy, confidentiality and security.<br>Audit trails. |
| C9 | Encrypted transmission. |
| C10 | Access and use on a need to know basis<br>Data user staff under a nondisclosure or confidentiality, or pledge of confidentiality<br>Privacy, confidentiality, and security training<br>Authentication measures<br>Special protection for remote electronic access to data |
| C11 | Minimising and specifying storage locations in advance, and organisational storage and access policies |
| C12 | Physical security measures. Protection from hazards such as fire and flood.<br>Routine video surveillance of premises in compliance with GDPR<br>Building design, alarms, locations control, control of visitors, photo ID, or coded card swipe, secure physical settings for computers and files |
| C13 | Audit trail |
| C14 | Data is securely disposed of after a specified retention period |
| C15 | Virus-checking and/or anti-malware programme |
| C16 | Sanctions for breach for privacy, confidentiality, or security, personal data breach and security incident protocol |
| C17 | Internal and external privacy reviews and audits<br>Data audit and monitoring mechanisms<br>See Appendix C |
| C18 | Internal and external privacy reviews and audits<br>Data audit and monitoring mechanisms<br>See Appendix C |
| C19 | |
| C20 | |
| C21 | Risk assessments have been conducted |
| C22 | Procedures to respond to public complaints or inquiries<br>Organisational transparency and public notification about personal data processing and anonymisation |
| C23 | DPO, data stewardship committees<br>Procedures to respond to public complaints or inquiries<br>Independent advisory or data stewardship committee<br>Organisational transparency and public notification about personal data processing and anonymisation<br>Designated senior individuals accountable for privacy and security |
| C24 | See also Appendix C |

| | | | |
|---|---|---|---|
| Close match | Partial match | Weak match | No match |

# 4. Supplemental Results

## 4.1. Targets and suspected matches

Intruders made 98,525 attempts and recorded 1,974 suspected matches (2.0%). Suspected match rates ranged from 0.46% to 8.07% across attack strategies (**Table 3**) and ranged from 1.06% (Belgium) to 2.65% (France) across countries (**Table 4**). **Table 5** lists the 30 regions with at least 10 suspected matches.

**Table 3.** Targeted individuals and suspected matches by attack strategy (P2S).

| Attack strategy | Targeted individuals | Suspected matches | Match rate |
|---|---|---|---|
| Events | 31,101 | 143 | 0.46% |
| Famous individuals | 125 | 0 | not analysed |
| Professionals | 1,768 | 6 | not analysed |
| Sporting events | 2,720 | 24 | 0.88% |
| Travellers searching home | 10,098 | 243 | 2.41% |
| Events automated | 7,029 | 567 | 8.07% |
| Events automated excluding self-search | 45,660 | 989 | 2.17% |
| **Total (P2S)** | **98,501** | **1,972** | **2.00%** |

Note. Strategies with fewer than 10 suspected matches are not analysed.

**Table 4.** Targeted individuals and suspected matches by country.

| Country | Targeted individuals | Suspected matches | Match rate |
|---|---|---|---|
| Germany | 30,848 | 598 | 1.94% |
| Spain | 25,094 | 484 | 1.93% |
| France | 19,048 | 505 | 2.65% |
| Poland | 10,018 | 143 | 1.43% |
| Ireland | 7,087 | 176 | 2.48% |
| Belgium | 6,426 | 68 | 1.06% |
| Not recorded | 4 | 0 | 0.00% |
| **Total** | **98,525** | **1,974** | **2.00%** |

Note. Both attack directions combined.

**Table 5.** Targeted individuals and suspected matches by NUTS 3 region.

| # | NUTS 3 | Country | Region | Targeted individuals and suspected matches | | | Confidence scores* | |
|---|---|---|---|---|---|---|---|---|
| | | | | Targeted individuals | Suspected matches | Match rate | n | Mean |
| 1 | ES511 | Spain | Barcelona | 13,061 | 184 | 1.41% | 42 | 2.64 |
| 2 | DE212 | Germany | München, Kreisfreie Stadt | 9,048 | 49 | 0.54% | 2 | 3.50 |
| 3 | FR101 | France | Paris | 5,960 | 233 | 3.91% | 5 | 3.40 |
| 4 | DE300 | Germany | Berlin | 5,110 | 364 | 7.12% | 13 | 3.15 |
| 5 | IE061 | Ireland | Dublin | 5,108 | 110 | 2.15% | 5 | 4.20 |
| 6 | FRK26 | France | Rhône | 4,958 | 168 | 3.39% | 1 | 3.00 |
| 7 | DE501 | Germany | Bremen, Kreisfreie Stadt | 4,739 | 18 | 0.38% | 17 | 4.12 |
| 8 | ES300 | Spain | Madrid | 4,489 | 121 | 2.70% | 18 | 3.94 |
| 9 | PL213 | Poland | Miasto Kraków | 4,099 | 34 | 0.83% | 12 | 1.83 |
| 10 | DE600 | Germany | Hamburg | 2,342 | 33 | 1.41% | 1 | 3.00 |
| 11 | BE242 | Belgium | Arr. Leuven | 1,766 | 19 | 1.08% | 0 | – |
| 12 | ES617 | Spain | Málaga | 1,639 | 83 | 5.06% | 2 | 4.50 |
| 13 | PL911 | Poland | Miasto Warszawa | 1,287 | 38 | 2.95% | 9 | 1.78 |
| 14 | ES614 | Spain | Granada | 1,207 | 26 | 2.15% | 0 | – |
| 15 | DE126 | Germany | Mannheim, Stadtkreis | 1,036 | 29 | 2.80% | 0 | – |
| 16 | BE234 | Belgium | Arr. Gent | 944 | 19 | 2.01% | 0 | – |
| 17 | FRG01 | France | Loire-Atlantique | 885 | 16 | 1.81% | 0 | – |
| 18 | PL514 | Poland | Miasto Wrocław | 801 | 10 | 1.25% | 1 | 3.00 |
| 19 | PL22A | Poland | Katowicki | 711 | 18 | 2.53% | 1 | 1.00 |
| 20 | FRJ23 | France | Haute-Garonne | 672 | 17 | 2.53% | 0 | – |
| 21 | IE042 | Ireland | West | 642 | 20 | 3.12% | 4 | 3.00 |
| 22 | PL633 | Poland | Trójmiejski | 593 | 20 | 3.37% | 0 | – |
| 23 | ES618 | Spain | Sevilla | 538 | 22 | 4.09% | 0 | – |
| 24 | FRJ13 | France | Hérault | 474 | 10 | 2.11% | 1 | 3.00 |
| 25 | PL711 | Poland | Miasto Łódź | 448 | 14 | 3.12% | 3 | 2.33 |
| 26 | DEA23 | Germany | Köln, Kreisfreie Stadt | 334 | 13 | 3.89% | 4 | 3.75 |
| 27 | DE929 | Germany | Region Hannover | 332 | 14 | 4.22% | 0 | – |
| 28 | IE052 | Ireland | South-East | 213 | 13 | 6.10% | 1 | 2.00 |
| 29 | DE134 | Germany | Ortenaukreis | 213 | 10 | 4.69% | 0 | – |
| 30 | BE212 | Belgium | Arr. Mechelen | 149 | 17 | 11.41% | 0 | – |

Note. Both attack directions combined. The 152 strata not shown are 150 regions with fewer than 10 suspected matches, one group of 28 attempts spanning several regions and 1 attempt with no region recorded; together they hold 24,727 attempts and 232 suspected matches. * Confidence scores are for manually reviewed suspected matches only: number of scores (n) and mean.

## 4.2. Confidence scores

The manually reviewed suspected matches had a mean confidence of 3.0, and 33% scored 4 or 5 (**Table 6**). Scores were lowest for sporting events (1.7) and highest for events (3.2) (**Table 7**), and country means ranged from 1.9 (Poland) to 3.5 (Germany) (**Table 8**). No attempt was made to calibrate Gemini confidence scores where there was no review.

**Table 6.** Confidence scores of manually reviewed suspected matches, in total, and by attack direction.

| Group | Confidence score (1 to 5): number of suspected matches | | | | | n | Mean |
|---|---|---|---|---|---|---|---|
| | 1 | 2 | 3 | 4 | 5 | | |
| **Total** | **31** | **19** | **67** | **40** | **18** | **175** | **2.97** |
| P2S | 29 | 19 | 67 | 40 | 18 | 173 | 2.99 |
| S2P | 2 | 0 | 0 | 0 | 0 | 2 | 1.00 |

**Table 7.** Confidence scores of manually reviewed suspected matches, by attack strategy.

| Attack strategy | Confidence score (1 to 5): number of suspected matches | | | | | n | Mean |
|---|---|---|---|---|---|---|---|
| | 1 | 2 | 3 | 4 | 5 | | |
| Events | 14 | 15 | 57 | 40 | 17 | 143 | 3.22 |
| Professionals | 1 | 0 | 4 | 0 | 1 | 6 | 3.00 |
| Sporting events | 14 | 4 | 6 | 0 | 0 | 24 | 1.67 |
| S2P | 2 | 0 | 0 | 0 | 0 | 2 | 1.00 |

Note. Only attack strategies with manually reviewed suspected matches are listed.

**Table 8.** Confidence scores of manually reviewed suspected matches, by country.

| Country | Confidence score (1 to 5): number of suspected matches | | | | | n | Mean |
|---|---|---|---|---|---|---|---|
| | 1 | 2 | 3 | 4 | 5 | | |
| Spain | 12 | 4 | 23 | 17 | 9 | 65 | 3.11 |
| Germany | 1 | 1 | 24 | 13 | 7 | 46 | 3.52 |
| Poland | 13 | 8 | 7 | 2 | 0 | 30 | 1.93 |
| France | 2 | 1 | 7 | 3 | 0 | 13 | 2.85 |
| Ireland | 0 | 3 | 3 | 3 | 2 | 11 | 3.36 |
| Belgium | 3 | 2 | 3 | 2 | 0 | 10 | 2.40 |

## 4.3. Time and effort

Time and effort for the different attacks are described in **Table 9** to **Table 11**.

**Table 9.** Testing hours by attack direction and for strategy and planning. Rounded to the nearest hour.

| Category | Hours |
|---|---|
| P2S | 153 |
| S2P | 32 |
| Strategy and planning | 98 |
| **Total** | **283** |

**Table 10.** Gemini API tokens consumed within RIDGE, by attack direction. Rounded to the nearest million.

| Direction | Input | Output (inc. thinking) | Total |
|---|---|---|---|
| P2S | 498,000,000 | 58,000,000 | 555,000,000 |
| S2P | 35,000,000 | 1,000,000 | 36,000,000 |
| **Total** | **533,000,000** | **58,000,000** | **591,000,000** |

**Table 11.** Testing hours per hit, by attack direction and overall.

| Category | Hours | Hits | Hours per hit |
|---|---|---|---|
| P2S (excluding strategy and planning) | 152.7 | 93 | 1.6 |
| S2P (excluding strategy and planning) | 31.9 | 0 | not defined |
| **Overall, including strategy and planning** | **282.6** | **93** | **3.0** |
| **Overall, excluding strategy and planning** | **184.6** | **93** | **2.0** |

## 4.4. Adjudication by Alphabet by country, NUTS 3 region and batch

**Table 12** and **Table 13** repeat the country and NUTS 3 results with the full set of metrics. Overall, the mini-session re-identification rate was 4.7% [3.2%, 10.5%] and the weighted re-identification rate (the mean, across submissions, of the proportion of each submission's mini-sessions that were correct) was 4.9% [3.9%, 6.6%]; counting unclear mini-sessions as correct gave 33.8% and 30.1% (**Table 12**). Alphabet reports a stratum only if it holds at least 10 submissions under the treatment used, so 7 of the 30 regions have no range and are shown only under unclear counted as correct. The 6,878 adjudicated mini-sessions comprised 6,609 unique mini-sessions, of which 186 (2.8%) were included in more than one submission (**Table 14**).

**Table 12.** Adjudication by Alphabet by country (P2S).

| Country | Submissions | Hits | Sub. re-ID rate | Mini-session re-ID rate | Weighted re-ID rate | Mini-session re-ID rate among hits | Mean correct IDs | Median correct IDs | Targeted individuals (P2S) | Targeted pop. re-ID rate | Pop. re-ID rate (regions) | Pop. re-ID rate (country) |
|---|---|---|---|---|---|---|---|---|---|---|---|---|
| **Unclear excluded [range]** | | | | | | | | | | | | |
| Germany | 470 [302, 597] | 30 | 6.4% [5.0%, 9.9%] | 5.4% [3.5%, 20.6%] | 5.4% [4.1%, 8.3%] | 78.1% [71.2%, 78.1%] | 4.03 | 1 | 30,846 | 0.097% | 0.000113% | 0.000036% |
| Spain | 415 [348, 483] | 21 | 5.1% [4.3%, 6.0%] | 5.0% [3.9%, 6.7%] | 4.7% [3.9%, 5.6%] | 90.7% [84.8%, 90.7%] | 1.86 | 1 | 25,090 | 0.084% | 0.000052% | 0.000043% |
| France | 426 [327, 505] | 28 | 6.6% [5.5%, 8.6%] | 4.2% [3.0%, 7.7%] | 5.7% [4.7%, 7.5%] | 66.7% [60.0%, 66.7%] | 1.50 | 1 | 19,041 | 0.147% | 0.000065% | 0.000041% |
| Poland | 107 [78, 142] | 6 | 5.6% [4.2%, 7.7%] | 3.2% [2.2%, 6.2%] | 3.9% [2.9%, 5.3%] | 52.9% | 1.50 | 1 | 10,013 | 0.060% | 0.000055% | 0.000016% |
| Ireland | 150 [108, 176] | 7 | 4.7% [4.0%, 6.5%] | 3.3% [2.5%, 5.9%] | 3.5% [2.6%, 4.8%] | 35.5% [31.4%, 35.5%] | 1.57 | 1 | 7,087 | 0.099% | 0.000129% | 0.000129% |
| Belgium | 60 [47, 68] | 1 | 1.7% [1.5%, 2.1%] | 0.7% [0.6%, 1.3%] | 1.7% [1.5%, 2.1%] | 100.0% | 1.00 | 1 | 6,424 | 0.016% | 0.000017% | 0.000008% |
| **All P2S** | **1,629 [1,211, 1,972]** | **93** | **5.7% [4.7%, 7.7%]** | **4.7% [3.2%, 10.5%]** | **4.9% [3.9%, 6.6%]** | **71.9% [65.8%, 71.9%]** | **2.40** | **1** | **98,501** | **0.094%** | **0.000071%** | **0.000036%** |
| **Unclear counted as correct** | | | | | | | | | | | | |
| Germany | 597 | 325 | 54.4% | 39.0% | 37.3% | 44.5% | 4.14 | 1 | 30,846 | 1.054% | 0.001229% | 0.000389% |
| Spain | 483 | 156 | 32.3% | 26.1% | 23.9% | 56.4% | 1.69 | 1 | 25,090 | 0.622% | 0.000387% | 0.000318% |
| France | 505 | 206 | 40.8% | 30.8% | 28.2% | 47.2% | 2.09 | 1 | 19,041 | 1.082% | 0.000480% | 0.000299% |
| Poland | 142 | 70 | 49.3% | 31.5% | 36.3% | 46.0% | 1.81 | 1 | 10,013 | 0.699% | 0.000646% | 0.000192% |
| Ireland | 176 | 75 | 42.6% | 28.8% | 27.6% | 44.0% | 1.72 | 1 | 7,087 | 1.058% | 0.001379% | 0.001379% |
| Belgium | 68 | 22 | 32.4% | 16.6% | 19.4% | 30.8% | 1.27 | 1 | 6,424 | 0.342% | 0.000370% | 0.000185% |
| **All P2S** | **1,972** | **854** | **43.3%** | **33.8%** | **30.1%** | **45.9%** | **2.72** | **1** | **98,501** | **0.867%** | **0.000648%** | **0.000334%** |

Note. Targeted pop. re-ID rate: hits divided by all P2S attempts logged for the country; excluded submissions stay in the denominator. Pop. re-ID rate (regions): hits divided by the residents of the NUTS 3 regions attacked in the country; (country): hits divided by the whole population of the country (Eurostat, 1 January 2025). The submission-level exclusion is applied before results are split by country. Under each treatment the country rows add up to one submission fewer than the overall row, because one submission is not allocated to any of the six countries.

**Table 13.** Adjudication by Alphabet by NUTS 3 region (P2S).

| NUTS 3 | Region | Submissions | Hits | Sub. re-ID rate | Mini-session re-ID rate | Weighted re-ID rate | Mini-session re-ID rate among hits | Mean correct IDs | Median correct IDs | Targeted individuals (P2S) | Targeted pop. re-ID rate | Pop. re-ID rate |
|---|---|---|---|---|---|---|---|---|---|---|---|---|
| **Unclear excluded [range]** | | | | | | | | | | | | |
| ES511 | Barcelona | 166 [144, 184] | 3 | 1.8% [1.6%, 2.1%] | 2.2% [1.8%, 2.6%] | 1.5% [1.3%, 1.7%] | 87.5% [77.8%, 87.5%] | 2.33 | 1 | 13,061 | 0.023% | 0.000050% |
| DE212 | München, Kreisfreie Stadt | 38 [29, 49] | 2 | 5.3% [4.1%, 6.9%] | 3.7% [2.2%, 5.1%] | 5.3% [4.1%, 6.9%] | 100.0% | 1.00 | 1 | 9,048 | 0.022% | 0.000133% |
| FR101 | Paris | 192 [122, 233] | 7 | 3.6% [3.0%, 5.7%] | 1.8% [1.2%, 5.1%] | 2.9% [2.3%, 4.5%] | 45.5% [38.5%, 45.5%] | 1.43 | 1 | 5,960 | 0.117% | 0.000339% |
| DE300 | Berlin | 298 [162, 364] | 10 | 3.4% [2.7%, 6.2%] | 0.9% [0.6%, 6.0%] | 2.4% [2.0%, 4.5%] | 40.5% [33.3%, 40.5%] | 1.70 | 1 | 5,110 | 0.196% | 0.000271% |
| IE061 | Dublin | 110 [79, 126] | 4 | 3.6% [3.2%, 5.1%] | 3.0% [2.2%, 5.3%] | 2.8% [2.5%, 4.0%] | 50.0% [46.7%, 50.0%] | 1.75 | 1 | 5,108 | 0.078% | 0.000257% |
| FRK26 | Rhône | 144 [126, 168] | 13 | 9.0% [7.7%, 10.3%] | 6.3% [5.2%, 9.1%] | 9.0% [7.3%, 10.3%] | 100.0% [90.0%, 100.0%] | 1.38 | 1 | 4,956 | 0.262% | 0.000674% |
| DE501 | Bremen, Kreisfreie Stadt | 15 [14, 18] | 5 | 33.3% [27.8%, 35.7%] | 66.7% [61.4%, 69.0%] | 31.3% [26.1%, 33.5%] | 95.1% | 15.60 | 11 | 4,739 | 0.106% | 0.000853% |
| ES300 | Madrid | 97 [68, 120] | 5 | 5.2% [4.2%, 7.4%] | 4.4% [3.2%, 8.7%] | 4.5% [3.1%, 6.4%] | 83.3% [71.4%, 83.3%] | 2.00 | 1 | 4,488 | 0.111% | 0.000070% |
| PL213 | Miasto Kraków | 27 [23, 34] | 3 | 11.1% [8.8%, 13.0%] | 4.3% [3.2%, 5.9%] | 6.2% [4.9%, 7.2%] | 33.3% | 1.00 | 1 | 4,099 | 0.073% | 0.000376% |
| DE600 | Hamburg | 29 [23, 33] | 3 | 10.3% [9.1%, 13.0%] | 9.8% [7.1%, 14.7%] | 10.3% [9.1%, 13.0%] | 100.0% | 1.67 | 2 | 2,342 | 0.128% | 0.000161% |
| BE242 | Arr. Leuven | 19 [16, 19] | 0 | 0.0% | 0.0% | 0.0% | – | – | – | 1,766 | 0.000% | 0.000000% |
| ES617 | Málaga | 74 [71, 83] | 4 | 5.4% [4.8%, 5.6%] | 3.8% [3.3%, 4.0%] | 5.4% [4.8%, 5.6%] | 100.0% | 1.00 | 1 | 1,638 | 0.244% | 0.000223% |
| PL911 | Miasto Warszawa | 29 [17, 38] | 1 | 3.4% [2.6%, 5.9%] | 2.6% [1.8%, 7.4%] | 1.7% [1.3%, 2.9%] | 50.0% | 2.00 | 2 | 1,287 | 0.078% | 0.000053% |

| NUTS 3 | Region | Submissions | Hits | Sub. re-ID rate | Mini-session re-ID rate | Weighted re-ID rate | Mini-session re-ID rate among hits | Mean correct IDs | Median correct IDs | Targeted individuals (P2S) | Targeted pop. re-ID rate | Pop. re-ID rate |
|---|---|---|---|---|---|---|---|---|---|---|---|---|
| ES614 | Granada | 19 [18, 26] | 4 | 21.1% [15.4%, 22.2%] | 21.4% [15.4%, 24.0%] | 21.1% [15.4%, 22.2%] | 100.0% | 1.50 | 1 | 1,207 | 0.331% | 0.000423% |
| DE126 | Mannheim, Stadtkreis | 11 [10, 29] | 0 | 0.0% | 0.0% | 0.0% | – | – | – | 1,036 | 0.000% | 0.000000% |
| BE234 | Arr. Gent | 15 [15, 19] | 0 | 0.0% | 0.0% | 0.0% | – | – | – | 943 | 0.000% | 0.000000% |
| FRG01 | Loire-Atlantique | 15 [11, 16] | 0 | 0.0% | 0.0% | 0.0% | – | – | – | 885 | 0.000% | 0.000000% |
| FRJ23 | Haute-Garonne | 13 [13, 17] | 2 | 15.4% [11.8%, 15.4%] | 8.7% [7.4%, 8.7%] | 11.5% [8.8%, 11.5%] | 66.7% | 1.00 | 1 | 672 | 0.298% | 0.000133% |
| IE042 | West | 16 [12, 20] | 1 | 6.2% [5.0%, 8.3%] | 7.7% [5.3%, 11.1%] | 6.2% [5.0%, 8.3%] | 100.0% | 2.00 | 2 | 642 | 0.156% | 0.000193% |
| ES618 | Sevilla | 21 [16, 22] | 0 | 0.0% | 0.0% | 0.0% | – | – | – | 538 | 0.000% | 0.000000% |
| PL711 | Miasto Łódź | 11 [10, 14] | 0 | 0.0% | 0.0% | 0.0% | – | – | – | 448 | 0.000% | 0.000000% |
| DEA23 | Köln, Kreisfreie Stadt | 11 [10, 12] | 1 | 9.1% [8.3%, 10.0%] | 5.9% [5.0%, 6.2%] | 9.1% [8.3%, 10.0%] | 100.0% | 1.00 | 1 | 334 | 0.299% | 0.000098% |
| BE212 | Arr. Mechelen | 15 [11, 17] | 1 | 6.7% [5.9%, 9.1%] | 2.6% [2.1%, 5.6%] | 6.7% [5.9%, 9.1%] | 100.0% | 1.00 | 1 | 149 | 0.671% | 0.000276% |
| **Unclear counted as correct** | | | | | | | | | | | | |
| ES511 | Barcelona | 184 | 43 | 23.4% | 21.3% | 16.9% | 61.3% | 1.95 | 1 | 13,061 | 0.329% | 0.000721% |
| DE212 | München, Kreisfreie Stadt | 49 | 22 | 44.9% | 41.6% | 35.0% | 71.2% | 1.68 | 1 | 9,048 | 0.243% | 0.001462% |
| FR101 | Paris | 233 | 118 | 50.6% | 35.4% | 32.2% | 44.7% | 2.52 | 1 | 5,960 | 1.980% | 0.005713% |
| DE300 | Berlin | 364 | 212 | 58.2% | 37.1% | 35.3% | 40.5% | 5.11 | 1 | 5,110 | 4.149% | 0.005753% |
| IE061 | Dublin | 126 | 50 | 39.7% | 27.1% | 24.8% | 44.0% | 1.70 | 1 | 5,108 | 0.979% | 0.003213% |
| FRK26 | Rhône | 168 | 55 | 32.7% | 23.3% | 24.9% | 47.9% | 1.47 | 1 | 4,956 | 1.110% | 0.002850% |
| DE501 | Bremen, Kreisfreie Stadt | 18 | 9 | 50.0% | 69.3% | 44.6% | 91.7% | 9.78 | 6 | 4,739 | 0.190% | 0.001535% |
| ES300 | Madrid | 120 | 57 | 47.5% | 32.2% | 31.9% | 47.7% | 1.79 | 1 | 4,488 | 1.270% | 0.000801% |
| PL213 | Miasto Kraków | 34 | 14 | 41.2% | 29.8% | 31.0% | 53.8% | 2.00 | 1 | 4,099 | 0.342% | 0.001753% |
| DE600 | Hamburg | 33 | 13 | 39.4% | 34.3% | 29.6% | 58.5% | 1.85 | 2 | 2,342 | 0.555% | 0.000698% |
| BE242 | Arr. Leuven | 19 | 3 | 15.8% | 5.9% | 4.2% | 17.6% | 1.00 | 1 | 1,766 | 0.170% | 0.000565% |
| ES617 | Málaga | 83 | 16 | 19.3% | 16.4% | 17.3% | 76.9% | 1.25 | 1 | 1,638 | 0.977% | 0.000893% |
| PL911 | Miasto Warszawa | 38 | 22 | 57.9% | 31.2% | 35.5% | 39.5% | 1.55 | 1 | 1,287 | 1.709% | 0.001173% |
| ES614 | Granada | 26 | 12 | 46.2% | 43.6% | 43.3% | 85.0% | 1.42 | 1 | 1,207 | 0.994% | 0.001269% |

| NUTS 3 | Region | Submissions | Hits | Sub. re-ID rate | Mini-session re-ID rate | Weighted re-ID rate | Mini-session re-ID rate among hits | Mean correct IDs | Median correct IDs | Targeted individuals (P2S) | Targeted pop. re-ID rate | Pop. re-ID rate |
|---|---|---|---|---|---|---|---|---|---|---|---|---|
| DE126 | Mannheim, Stadtkreis | 29 | 19 | 65.5% | 63.2% | 63.2% | 92.3% | 1.26 | 1 | 1,036 | 1.834% | 0.005974% |
| BE234 | Arr. Gent | 19 | 4 | 21.1% | 26.1% | 21.1% | 100.0% | 1.50 | 1 | 943 | 0.424% | 0.000685% |
| FRG01 | Loire-Atlantique | 16 | 5 | 31.2% | 8.8% | 16.7% | 45.5% | 1.00 | 1 | 885 | 0.565% | 0.000331% |
| PL514 | Miasto Wrocław | 10 | 3 | 30.0% | 27.3% | 30.0% | 100.0% | 1.00 | 1 | 801 | 0.375% | 0.000446% |
| PL22A | Katowicki | 18 | 10 | 55.6% | 26.4% | 39.9% | 30.3% | 2.30 | 1 | 711 | 1.406% | 0.001487% |
| FRJ23 | Haute-Garonne | 17 | 6 | 35.3% | 22.2% | 32.4% | 85.7% | 1.00 | 1 | 672 | 0.893% | 0.000399% |
| IE042 | West | 20 | 9 | 45.0% | 36.8% | 33.3% | 63.6% | 1.56 | 1 | 642 | 1.402% | 0.001741% |
| PL633 | Trójmiejski | 20 | 12 | 60.0% | 54.3% | 48.6% | 71.4% | 2.08 | 1 | 593 | 2.024% | 0.001614% |
| ES618 | Sevilla | 22 | 6 | 27.3% | 20.6% | 14.8% | 46.7% | 1.17 | 1 | 538 | 1.115% | 0.000303% |
| FRJ13 | Hérault | 10 | 4 | 40.0% | 44.4% | 35.7% | 72.7% | 2.00 | 1 | 474 | 0.844% | 0.000319% |
| PL711 | Miasto Łódź | 14 | 4 | 28.6% | 11.4% | 22.9% | 50.0% | 1.00 | 1 | 448 | 0.893% | 0.000624% |
| DEA23 | Köln, Kreisfreie Stadt | 12 | 3 | 25.0% | 20.0% | 20.8% | 80.0% | 1.33 | 1 | 334 | 0.898% | 0.000293% |
| DE929 | Region Hannover | 14 | 7 | 50.0% | 37.0% | 35.7% | 62.5% | 1.43 | 1 | 332 | 2.108% | 0.000614% |
| DE134 | Ortenaukreis | 10 | 4 | 40.0% | 25.0% | 23.3% | 45.5% | 1.25 | 1 | 213 | 1.878% | 0.000899% |
| IE052 | South-East | 13 | 7 | 53.8% | 48.0% | 44.0% | 70.6% | 1.71 | 1 | 213 | 3.286% | 0.001443% |
| BE212 | Arr. Mechelen | 17 | 7 | 41.2% | 19.1% | 24.6% | 30.0% | 1.29 | 1 | 149 | 4.698% | 0.001931% |

Note. Regions without a range because Alphabet does not report them under every exclusion treatment (fewer than 10 submissions): PL514, PL22A, PL633, FRJ13, DE929, DE134, IE052. Dashes mean no hits. NUTS 3 location comes from each mini-session, and the submission-level exclusion is applied before splitting. One Irish pull spanned three regions (IE041, IE053 and IE061), so at most 19 submissions can count in more than one region, as a hit in one and a miss in another.

**Table 14.** Mini-sessions by number of times they appear, as reported by Alphabet.

| Times the identifier appears | Mini-sessions |
|---|---|
| 1 | 6,423 |
| 2 | 143 |
| 3 | 21 |
| 4 | 11 |
| 5 | 5 |
| 6 | 5 |
| 7 | 1 |
| **Distinct identifiers** | **6,609** |
| **Two or more times** | **186** |
| **Total appearances (identifiers adjudicated)** | **6,878** |

Note. The appearances add up to the mini-sessions adjudicated, and the mini-session metrics count each appearance.

## 4.5. Sensitivity to unclear outcomes

The 93 hits do not change unless unclear mini-sessions are counted as correct, so the submission re-identification rate of 5.7% [4.7%, 7.7%] differs across the exclusion strategies only through the denominators (**Table 16**). Counting unclear mini-sessions as correct gave 854 hits, a submission re-identification rate of 43.3% and a targeted population re-identification rate of 0.867%. Hours per hit for P2S fall from 1.6 to 0.2 under that treatment (**Table 15**). By attack strategy, the submission re-identification rate was 12.6% [11.0%, 17.8%] for events, 4.8% [4.1%, 8.6%] for events automated and 6.8% [5.5%, 8.0%] for events automated excluding self-search, with no hits among travellers searching home or sporting-event and professional submissions (**Table 18**).

**Table 15.** Testing hours per hit with unclear outcomes excluded and counted as correct.

| Category | Hours | Hits | | Hours per hit | |
|---|---|---|---|---|---|
| | | **Unclear excluded** | **Unclear counted as correct** | **Unclear excluded** | **Unclear counted as correct** |
| P2S (excluding strategy and planning) | 152.7 | 93 | 854 | 1.6 | 0.2 |
| S2P (excluding strategy and planning) | 31.9 | 0 | 0 | not defined | not defined |
| **Overall, including strategy and planning** | **282.6** | **93** | **854** | **3.0** | **0.3** |
| **Overall, excluding strategy and planning** | **184.6** | **93** | **854** | **2.0** | **0.2** |

**Table 16.** Overall metrics (P2S).

| **Metric** | **Unclear excluded [range]** | **Unclear counted as correct** |
|---|---|---|
| Submissions | 1,629 [1,211, 1,972] | 1,972 |
| Hits | 93 | 854 |
| **Submission re-identification rate** | **5.7% [4.7%, 7.7%]** | **43.3%** |
| Mini-sessions | 4,778 [2,129, 6,878] | 6,878 |
| Correct mini-sessions | 223 | 2,323 |
| Mini-session re-identification rate | 4.7% [3.2%, 10.5%] | 33.8% |
| Weighted re-identification rate | 4.9% [3.9%, 6.6%] | 30.1% |
| Mini-session re-identification rate among re-identifications | 71.9% [65.8%, 71.9%] | 45.9% |
| Mean correct identifiers among re-identifications | 2.40 | 2.72 |
| Median correct identifiers among re-identifications | 1 | 1 |
| Targeted individuals (P2S) | 98,501 | 98,501 |
| **Targeted population re-identification rate** | **0.094%** | **0.867%** |
| Population re-identification rate, attacked NUTS 3 regions (131,821,205 residents) | 0.000071% | 0.000648% |
| Population re-identification rate, whole countries attacked (255,409,305 residents) | 0.000036% | 0.000334% |
| Population re-identification rate, EEA (456,671,641 residents) | 0.000020% | 0.000187% |

**Table 17.** Metrics by candidate match review strategy (P2S).

| **Metric** | **Manual review** | **No review** |
|---|---|---|
| **Unclear excluded [range]** | | |
| Submissions | 141 [100, 173] | 1,488 [1,111, 1,799] |
| Hits | 16 | 77 |
| **Submission re-identification rate** | **11.3% [9.2%, 16.0%]** | **5.2% [4.3%, 6.9%]** |
| Mini-session re-identification rate | 22.3% [17.1%, 34.1%] | 2.9% [2.0%, 6.7%] |
| Weighted re-identification rate | 9.1% [7.3%, 12.9%] | 4.5% [3.6%, 6.0%] |
| Mini-session re-identification rate among re-identifications | 87.6% [86.1%, 87.6%] | 62.9% [55.4%, 62.9%] |
| Mean correct identifiers among re-identifications | 6.19 | 1.61 |
| Median correct identifiers among re-identifications | 1 | 1 |
| **Targeted population re-identification rate** | **0.045%** | **0.122%** |
| Population re-identification rate (attacked regions) | 0.000017% | 0.000063% |
| Population re-identification rate (countries attacked) | 0.000006% | 0.000030% |
| **Unclear counted as correct** | | |
| Submissions | 173 | 1,799 |
| Hits | 89 | 765 |
| Submission re-identification rate | 51.4% | 42.5% |
| Mini-session re-identification rate | 40.5% | 33.2% |
| Weighted re-identification rate | 35.5% | 29.6% |
| Mini-session re-identification rate among re-identifications | 58.4% | 44.8% |
| Mean correct identifiers among re-identifications | 2.63 | 2.73 |
| Median correct identifiers among re-identifications | 1 | 1 |
| Targeted population re-identification rate | 0.251% | 1.215% |
| Population re-identification rate (attacked regions) | 0.000096% | 0.000627% |
| Population re-identification rate (countries attacked) | 0.000035% | 0.000300% |

**Table 18.** Metrics by attack strategy (P2S).

| Metric | Events | Sporting events and professionals | Events automated | Events automated excluding self-search | Travellers searching home |
|---|---|---|---|---|---|
| **Unclear excluded [range]** | | | | | |
| Submissions | 127 [90, 145] | 14 [10, 28] | 482 [267, 567] | 797 [671, 989] | 209 [173, 243] |
| Hits | 16 | 0 | 23 | 54 | 0 |
| Submission re-identification rate | 12.6% [11.0%, 17.8%] | 0.0% | 4.8% [4.1%, 8.6%] | 6.8% [5.5%, 8.0%] | 0.0% |
| Mini-session re-identification rate | 23.7% [18.8%, 36.1%] | 0.0% | 1.6% [1.1%, 8.0%] | 5.7% [4.2%, 8.1%] | 0.0% |
| Weighted re-identification rate | 10.1% [8.7%, 14.3%] | 0.0% | 3.2% [2.5%, 5.7%] | 6.4% [5.0%, 7.7%] | 0.0% |
| Mini-session re-identification rate among re-identifications | 87.6% [86.1%, 87.6%] | – | 38.8% [32.5%, 38.8%] | 89.4% [83.2%, 89.4%] | – |
| Mean correct identifiers among re-identifications | 6.19 | – | 1.74 | 1.56 | – |
| Median correct identifiers among re-identifications | 1 | – | 1 | 1 | – |
| Targeted population re-identification rate | 0.051% | 0.000% | 0.327% | 0.118% | 0.000% |
| Population re-identification rate (attacked regions) | 0.000020% | 0.000000% | 0.000035% | 0.000048% | 0.000000% |
| Population re-identification rate (countries attacked) | 0.000006% | 0.000000% | 0.000009% | 0.000021% | 0.000000% |
| **Unclear counted as correct** | | | | | |
| Submissions | 145 | 28 | 567 | 989 | 243 |
| Hits | 71 | 18 | 323 | 372 | 70 |
| Submission re-identification rate | 49.0% | 64.3% | 57.0% | 37.6% | 28.8% |
| Mini-session re-identification rate | 39.5% | 50.0% | 36.2% | 30.5% | 21.0% |
| Weighted re-identification rate | 31.4% | 56.8% | 33.1% | 29.8% | 20.6% |
| Mini-session re-identification rate among re-identifications | 57.0% | 72.2% | 40.4% | 58.3% | 49.3% |
| Mean correct identifiers among re-identifications | 2.93 | 1.44 | 4.25 | 1.63 | 1.56 |
| Median correct identifiers among re-identifications | 1 | 1 | 1 | 1 | 1 |
| Targeted population re-identification rate | 0.227% | 0.416% | 4.595% | 0.815% | 0.693% |
| Population re-identification rate (attacked regions) | 0.000090% | 0.000033% | 0.000485% | 0.000333% | 0.000197% |
| Population re-identification rate (countries attacked) | 0.000028% | 0.000007% | 0.000126% | 0.000146% | 0.000027% |

Note. Professionals with sporting events were combined. A dash means no hits.

**Table 19.** Submission and targeted population re-identification rates by attack strategy.

| Attack strategy | Targeted individuals | Hits | Submissions | Submission re-ID rate | Targeted population re-ID rate |
|---|---|---|---|---|---|
| Events | 31,264 | 16 | 127 [90, 145] | 12.6% [11.0%, 17.8%] | 0.051% |
| Events automated | 7,029 | 23 | 482 [267, 567] | 4.8% [4.1%, 8.6%] | 0.327% |
| Events automated excluding self-search | 45,660 | 54 | 797 [671, 989] | 6.8% [5.5%, 8.0%] | 0.118% |
| Travellers searching home | 10,098 | 0 | 209 [173, 243] | 0.0% | 0.000% |
| Sporting events and professionals | 4,325 | 0 | 14 [10, 28] | 0.0% | 0.000% |

Note. Professionals were combined with sporting events. Famous individuals (125 attempts, no suspected matches) are not shown, so attempts by strategy add up to 125 fewer than all attempts. The 198 doctor-portal attempts, which had no manual review and no suspected matches, are counted under professionals.